\documentclass[12pt]{article}
\pdfoutput=1
\usepackage{jheppub}
\usepackage{amsfonts, amsthm}
\usepackage[english]{babel}
\usepackage[utf8]{inputenc}
\usepackage{slashed}
\usepackage{mathrsfs}
\usepackage{amssymb}
\usepackage{color}
\hypersetup{unicode}
\usepackage{natbib}
\usepackage{lipsum}
\usepackage{breqn}
\usepackage{physics}
\usepackage{youngtab}
\newcommand{\eq}{\begin{equation}}
\newcommand{\feq}{\end{equation}}
\newcommand{\eqn}{\begin{eqnarray}}
\newcommand{\feqn}{\end{eqnarray}}

\def\Tr{{\mathrm{Tr}}}
\def\beq{\begin{align}}
\def\eeq{\end{align}}
\newcommand{\bi}{\begin{itemize}}
\newcommand{\ei}{\end{itemize}}
\newcommand{\ben}{\begin{enumerate}}
\newcommand{\een}{\end{enumerate}}
\providecommand{\bea}{\begin{eqnarray}}
\providecommand{\eea}{\end{eqnarray}}
\newcommand{\be}{\begin{equation}}
\newcommand{\ee}{\end{equation}}

\def\Tr{{\mathrm{Tr}}}

\definecolor{applegreen}{rgb}{0.55, 0.71, 0.0}

\newcommand{\Dp}{{\mathrm D}}
\newcommand{\Op}{{\mathrm O}}

\newcommand{\cA}{\mathcal{A}}
\newcommand{\cF}{\mathcal{F}}
\newcommand{\cL}{\mathcal{L}}
\newcommand{\cM}{\mathcal{M}}
\newcommand{\cN}{\mathcal{N}}
\newcommand{\cO}{\mathcal{O}}
\newcommand{\cR}{\mathcal{R}}
\newcommand{\cV}{\mathcal{V}}
\renewcommand{\Tr}[1]{\mathrm{Tr}\left( #1 \right)}

\newcommand{\nv}{\mathfrak{N}}

\title{Open Strings in IIB Orientifold Reductions}

\author[a]{Juan Ram\'on Balaguer,}
\author[b,c]{Giuseppe Dibitetto,}
\author[d]{Jose J. Fern\'andez-Melgarejo,}
\author[b,c]{Alejandro Ruip\'erez}

\affiliation[a]{Departamento de F\'isica, Universidad de Murcia, Campus de Espinardo, E-30100 Murcia, Spain}
\affiliation[b]{Dipartimento di Fisica, Universit\`a di Roma ``Tor Vergata", Via della Ricerca Scientifica 1, 00133, Roma, Italy}
\affiliation[c]{INFN, Sezione di Roma2, Via della Ricerca Scientifica 1, 00133, Roma, Italy}
\affiliation[d]{Departamento de Electromagnetismo y Electr\'onica, Universidad de Murcia, Campus de Espinardo, E-30100 Murcia, Spain}

\emailAdd{juanramon.balaguer@um.es}
\emailAdd{giuseppe.dibitetto@roma2.infn.it}
\emailAdd{melgarejo@um.es}
\emailAdd{alejandro.ruiperez@roma2.infn.it}

\date{\today}

\abstract{We consider type IIB compactifications on a general 4D group manifold with different types of possible spacetime filling O-planes and the corresponding D-branes parallel to them. Once fluxes allowed by the associated orientifold projection are included, a 6D $\mathcal{N}=(1,1)$ gauged supergravity is obtained. In this paper we show how the consistent coupling to dynamical open strings living on the spacetime filling D-branes may be captured by the inclusion of extra vector multiplets and extra embedding tensor deformations on the gauged supergravity side. As a result, the quadratic constraints on the embedding tensor consistently reproduce the source corrected 10D Bianchi identities. Furthermore, the field strength modifications induced by the open string sector could potentially be understood as U-dual versions of the Green-Schwarz terms. Finally, the entire scalar potential of the theory exactly matches the one obtained from reduction of the bulk action plus the source contributions.}


\begin{document}
\maketitle
\flushbottom

\section{Introduction}
Extracting consistent low energy effective descriptions from string theory is one the main challenges of theoretical high energy physics. Typically this procedure involves dimensional reduction and (partial) supersymmetry breaking. Depending on the mechanism used in order to realize them, a plethora of viable lower dimensional models arises, with varying amounts of supersymmetry in different dimensions. Low energy effective models obtained in this way are by construction UV consistent and belong to the string landscape.

On the other hand, by adopting a \emph{bottom-up} approach instead, one could study  different lower dimensional (supersymmetric) constructions and assess whether or not they can be consistently coupled with quantum gravity in a UV regime. This is the perspective promoted by the so-called Swampland Program \cite{Vafa:2005ui,Ooguri:2006in}, which aims at identifying a set of consistency requirements that any effective theory must comply with, in order for it to admit a UV completion.

By restricting oneself to theories enjoying extended supersymmetry, the range of possibilities gets drastically reduced, up to the extent that UV consistency requirements in some instances may be even exhaustively analyzed. This certainly applies to the case of maximal supersymmetry, \emph{i.e.} $32$ supercharges. In 10D, the only consistent maximal supergravities are type IIA and type IIB supergravities and they exactly coincide with the low energy limits of the corresponding superstring theories, respectively. This may be viewed as a prime manifestation of \emph{string universality}.

Now, still within 10D one may consider theories with half-maximal supersymmetry. In such a situation, the aforementioned universality principle was verified in \cite{Adams:2010zy} by showing that the only UV consistent half-maximal theories are $\mathcal{N}=1$ supergravities with gauge groups given by either $\mathrm{SO}(32)$ or $\mathrm{E}_8\times\mathrm{E}_8$. Those are indeed the only gauge symmetries that may be ever obtained by considering the low energy limits of heterotic or type I superstring theories.

In the last few decades we have learned a number of things concerning string theories with $16$ supercharges and this allowed us to address the string universality issue in dimension lower than $10$. By now we can consider it to be fully verified up to dimension $8$ \cite{Cvetic:2020kuw,Cvetic:2021sjm,Bedroya:2021fbu}. Besides, there have been recent developments even in dimension $7$ and $6$, the latter both with $(2,0)$ \cite{Taylor:2019ots} and $(1,1)$ \cite{Fraiman:2022aik} supersymmetry, as well as some preliminary studies on $D<6$ cases \cite{Fraiman:2018ebo,Font:2020rsk,Font:2021uyw}.

Our present work is to be placed within such a context, from which it draws its main motivations. We aim at building a bridge between top-down string theory constructions yielding 6D theories with $(1,1)$ supersymmetry and the corresponding (gauged) supergravities, which are classified by means of bottom-up based organizing principles\footnote{Some related works in 4D exist for $\cN=4$ \cite{Angelantonj:2003rq,Angelantonj:2003up,Roest:2009dq}, $\cN=2$ \cite{Angelantonj:2003zx}, and more recently, \cite{Andriot:2022bnb}.}. 
In more concrete terms, the stringy setup's relevant here are compactifications of type I/heterotic strings on $\mathbb{T}^4$, as well as orientifold reductions of type IIA/IIB on $\mathbb{T}^4$, or M theory on $\mathbb{T}^5$. Our interest towards $(1,1)$ supergravity rather than for the $(2,0)$ one is due to the fact that none of the orientifold projections respecting chiral extended 6D supersymmetry allows to turn on fluxes. This is, on the other hand, consistent with the statement that $(2,0)$ supergravities do not admit any consistent embedding tensor deformations. Conversely in the non-chiral $(1,1)$ case, we find a wide range of flux compactifications.

In \cite{Dibitetto:2019odu}, an analysis of this sort was already presented and all the cases consistent with 6D Lorentz symmetry and $(1,1)$ supersymmetry were discussed. In each single setup the dictionary was obtained between 10D (11D) fields \& fluxes on the one side, and 6D fields \& deformations on the other side. The approach used mimics that of \cite{Dibitetto:2010rg,Dibitetto:2011gm} designed for orientifold reductions down to 4D. Focusing on compactifications over 4d twisted tori, a vacua scan performed with the aid of the 6D gauged supergravity description showed the existence of a wide landscape of Minkowski (Mkw) vacua, but no maximally symmetric backgrounds with non-vanishing cosmological constant appeared.

The aim of this paper is to extend the analysis carried out in \cite{Dibitetto:2019odu} to include open string effects such as dynamical brane position moduli and Wilson lines wrapped in internal space, as well as non-Abelian brane gauge groups and non-trivial associated Yang-Mills (YM) flux. While these ingredients are difficult to take into account from a top-down perspective, we show that these are straightforwardly handled from a bottom-up viewpoint, just at the price of including extra vector multiplets within the associated gauged supergravity description. The reason for this is that Lagrangians of half-maximal gauged supergravities (see \emph{e.g.} \cite{Schon:2006kz} for the 4D \& 5D cases) are fully determined for a given choice of embedding tensor \cite{deWit:2002vt}, simply out of imposing consistency and supersymmetry.

At a technical level, the work done in this paper proves the equivalence between the effective scalar potential obtained from type IIB orientifold reductions and that of a suitable gauged supergravity, with the appropriate amount of vector multiplets accounting for both closed and open string excitations. The highly non-trivial result is a full matching between the scalar potential obtained from gauged supergravity and the one arising from reduction of the bulk action plus the contributions coming from the effective actions of the spacetime filling sources. This matching works even in presence of open string effects such as non-Abelian brane gauge groups and non-vanishing YM internal flux. It is worth remarking that such competing effects between closed and open string sectors in some sense require working at \emph{finite} $\alpha'$. It still remains to be understood whether this set of $\alpha'$ effects is also physically reliable, besides being mathematically consistent.

The paper is organized as follows. In Sec.~\ref{section:Oplanes} we review some salient features of O$p$/D$p$ systems, the associated light dof's, possible gauge groups and consistency requirements. In Sec. \ref{section:6D_sugra} we spell out the embedding tensor formulation of 6D $\mathcal{N}=(1,1)$ gauged supergravities coupled to an arbitrary number of vector multiplets. In Sec.~\ref{section:O5} we analyze the case of IIB reductions including spacetime filling O$5$/D$5$ sources and work out the dictionary between the 6D supergravity side and the type IIB side. 
A parallel analysis is then carried out in Sec.~\ref{section:O7} for O$7$/D$7$ sources and later in Sec.~\ref{section:O9} for O$9$/D$9$, \emph{i.e.} type I reductions. One of the key results of the paper is the discovery of bulk field strength modifications sourced by the open string vector fields, just like in the heterotic case, where this was due to the Green-Schwarz (GS) mechanism \cite{Green:1984sg}. Indeed, the modifications derived here could be heuristically understood as U-dual versions of GS terms. Finally, our appendices contain technical support material concerning type IIB supergravity and non-Abelian brane actions, as well as reductions thereof.

\section{General Aspects: O$p$/D$p$ Systems \& Open Strings}
\label{section:Oplanes}
Within the string theory spectrum D$p$-branes appear as higher dimensional spacetime defects representing dynamical boundary conditions for open strings. Such extended objects admit a low energy description in terms of supergravity black brane solutions.\footnote{See \cite{Ortin:2015hya} for an exhaustive review of such solutions.} In particular, a D$p$-brane has a positive tension 
\begin{equation}
T_{\text{D}p} \ = \ 2\pi\ell_s^{-(p+1)} \ ,
\end{equation}
and carries a positive charge $\mu_{\text{D}p}=T_{\text{D}p}$ w.r.t. a RR $(p+1)$-form field. The corresponding anti-brane $\overline{\textrm{D}p}$ will have the same tension, but carry opposite charge. Since the associated supersymmetry projectors are mutually orthogonal, brane configurations involving both D$p$'s \& $\overline{\textrm{D}p}$'s at the same time will necessarily be non-supersymmetric. 

Each D$p$-brane has a massless vector multiplet associated with the light open string state attached to it. Its low energy description is given by $\textrm{U}(1)$ maximal SYM in $(p+1)$ dimensions. Now, a set of $N$ D$p$-branes which are kept separated from one another at finite distance describes an Abelian $\textrm{U}(1)^N$ gauge theory. However though, in the limit where these are made to collide together to form a brane stack, the system undergoes a gauge symmetry enhancement to the non-Abelian gauge group $\textrm{U}(N)$. In this case, the $N^2$ generators of $\textrm{U}(N)$ are in one-to-one correspondence with light strings having each extremum on any D-brane within the stack. We refer to Appendix \ref{app:iib-DBI} for more details concerning non-Abelian brane actions and their relation to non-commutative geometry.

Besides ordinary D$p$-branes, more exotic objects are present in the spectrum, \emph{i.e.} orientifold planes. These objects are the loci of fixed points of a given orientifold $\mathbb{Z}_2$ action $\Omega_{\text{O}p}$, which may be defined through
\begin{equation}
\Omega_{\text{O}p} \ = \ \Omega \ \sigma_{\mathrm{O}p}  \ \sigma_{F_L} \ ,
\end{equation}
where $\Omega$ is the worldsheet parity acting on the closed string bulk fields as
\begin{equation}
\begin{array}{lll}
\Omega: \quad\qquad & 
\begin{array}{l}
G_{MN} \ \rightarrow \ G_{MN}  \ ,\\
B_{MN}  \ \rightarrow \ -B_{MN} \ ,\\
\Phi  \ \rightarrow \ \Phi \ ,
\end{array} \quad & \quad 
C_{(k)} \ \rightarrow \ (-1)^{q+r} C_{(k)} \ ,
\end{array}
\end{equation}
where $k= 2q+r$, with $r=0$ (type IIB), or $r=1$ (type IIA).
The second $\mathbb{Z}_2$ factor $\sigma_{\mathrm{O}p}$ is a spacetime involution flipping the sign of all transverse coordinates
\be
\mathrm{O}p \ : \quad
\underbrace{\times \ \cdots \ \times}_{(p+1)\textrm{D worldvolume}} \ | \ \underbrace{- \ \cdots \ -}_{\textrm{transverse }y^{i}} \ , \qquad 
\sigma_{\mathrm{O}p} \ : \ y^{i} \ \longrightarrow \ -y^{i} \ .
\notag
\ee
Finally, $\sigma_{F_L}$ involves the so-called fermionic number and is given by
\begin{equation}
\sigma_{F_L} \ = \ \left\{
\begin{array}{clcc}
(-1)^{F_L} & , & \quad p=2,3 \ \mathrm{mod} \ 4 & , \\
1 & , & \quad p=0,1 \ \mathrm{mod} \ 4 & .
\end{array}\right.
\end{equation}
It turns out that there exist two different types of O$p$-planes preserving the same supersymmetries as a stack of D$p$-branes parallel to them. 
These are conventionally denoted by O$p^+$ \& O$p^-$ and their tension $T_{{{\rm O}p}^{\pm}}$ and charge $\mu_{{{\rm O}p}^{\pm}}$ satisfy the following formula
\begin{equation}
T_{\text{O}p^{\pm}} \ = \ \mu_{\text{O}p^{\pm}} \ = \ \pm\, 2^{p-4} \,T_{\text{D}p} \ ,
\end{equation}
which in particular implies that fractional orientifold charges are allowed for $p<4$ \cite{Bergman:2001rp}. Despite the fact that O$p$-planes carry tension and charge, they appear to be completely rigid objects, at least at a perturbative level. 

\begin{figure}[h!]
	\centering
	\includegraphics[scale=0.5]{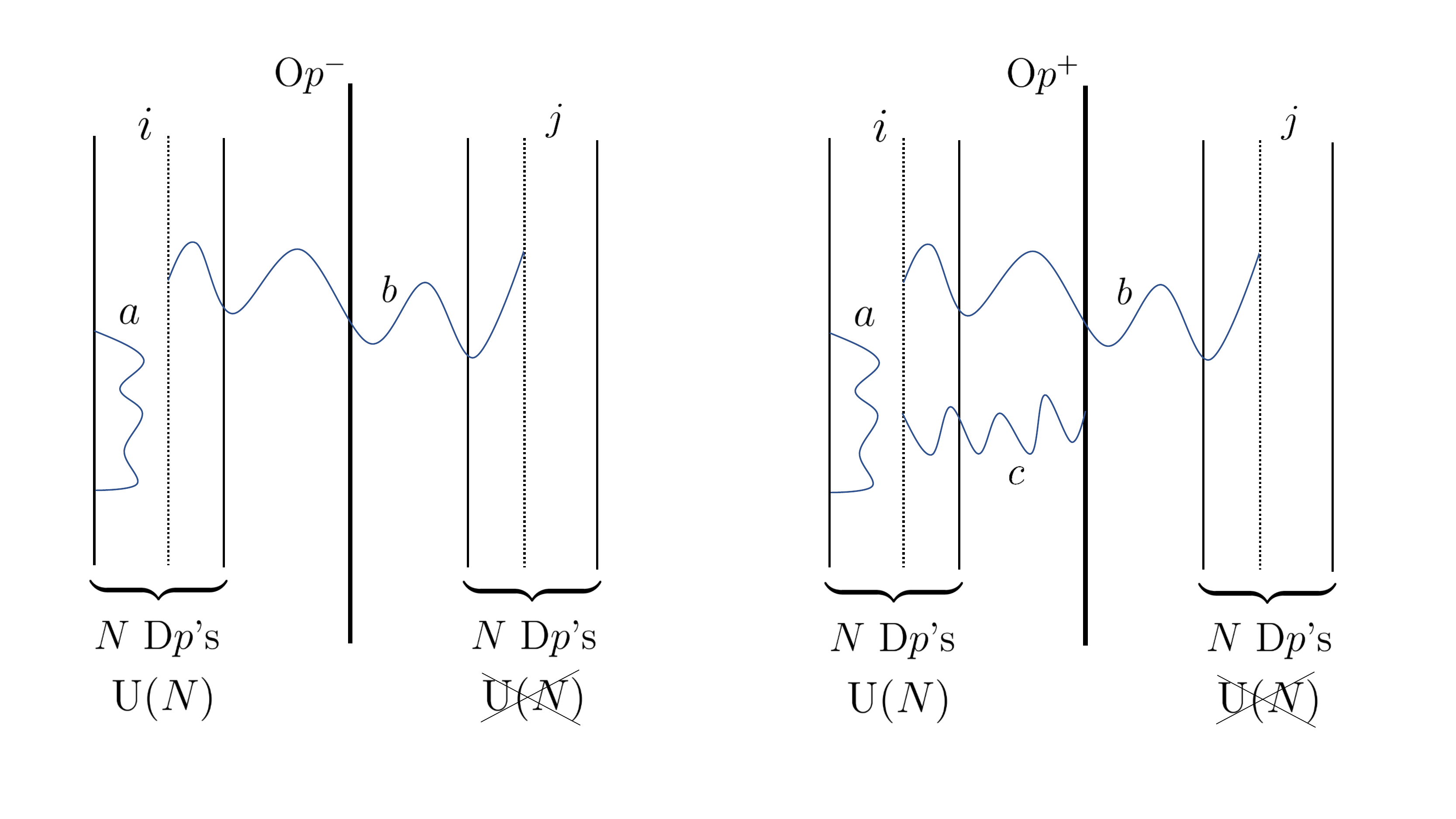}
	\caption{\it \emph{(Left)} In the presence of an $\mathrm{O}p^-$, a stack of $N$ coincident $\mathrm{D}p$-branes realizes an $\mathrm{SO}(2N)$ gauge group. Its $N(2N-1)$ light dof's can be understood as all open strings with both extrema on one side of the O-plane (type a, $N^2$ states), plus those with one extremum on each side, with the Chan-Paton rule that $i\neq j$ (type b, $N(N-1)$ states). \emph{(Right)} In the presence of an $\mathrm{O}p^+$, we still have open strings of type a \& b (these are now $N^2$ states due to the absence of the $i\neq j$ rule), and in addition we find strings connecting each D-brane to the O-plane (type c, $N$ states). This yields a total of $N(2N+1)$ light states realizing $\mathrm{USp}(2N)$.}
	\label{fig:SOvsUSp}
\end{figure}

When considering a system made out of D$p$-branes and parallel O$p$-planes, in order to fully specify the dynamics, we also need to spell out the orientifold action on the open string states living on the D-branes. This is done by identifying its action on the open string Chan-Paton factors $\lambda$. For $N$ D$p$'s and one O$p$ \cite{Gimon:1996rq}, this reads
\begin{equation}
\lambda \ \overset{\Omega}{\rightarrow} M[\Omega]^{-1}\lambda^{\text{T}} M[\Omega] \ , \qquad \textrm{with} \quad 
 M[\Omega] \ = \ \left\{
\begin{array}{cl}
\mathbb{I}_{2N} & \textrm{, for O}p^-,\\
\mathbb{J}_{2N} \equiv \left(\begin{array}{cc}\mathbb{O}_{N} & i\,\mathbb{I}_{N} \\ -i\,\mathbb{I}_{N} & \mathbb{O}_{N}\end{array}\right) & \textrm{, for O}p^+.
\end{array}\right.
\notag
\end{equation}
The above difference in the orientifold action at the level of the Chan-Paton factors results in different open string SYM gauge groups in presence of an O$p^+$, or an O$p^-$. In the former case we have an $\mathrm{USp}(2N)$ group, while in the latter we have $\mathrm{SO}(2N)$ instead. The corresponding conceptual picture in these two situations can be found in Figure \ref{fig:SOvsUSp}.

For the most general system made out of D$p$-branes \& O$p$-planes in the absence of fluxes and in flat space, the following tadpole cancellation condition is required for UV-finiteness of the corresponding quantum description
\begin{equation}
N_{\text{D}p}\mu_{\text{D}p} \ + \ N_{\overline{\text{D}p}}\mu_{\overline{\text{D}p}} \ + \ N_{\text{O}p^+}\mu_{\text{O}p^+} \ + \ N_{\text{O}p^-}\mu_{\text{O}p^-} \ \overset{!}{=} \ 0 \ ,
\end{equation}
which can be written as
\begin{equation}
\label{Tadpole_DpDpbar}
\left(N_{\text{D}p} \, - \, N_{\overline{\text{D}p}}\right)\  \overset{!}{=}  \ 2^{p-4}\left( N_{\text{O}p^-} \, - \, N_{\text{O}p^+}\right)\ ,
\end{equation}
where $N_{\text{D}p}=2N$ accounts for the imagine branes as well.
It is crucial to remember that the above constraint originates from demanding that string amplitudes be free of divergences and it refers to amplitudes calculated in flat space and in the absence of fluxes. In our work we will be considering more involved situations where the background fluxes may contribute in several ways to the tadpole constraints for the corresponding spacetime filling sources. We will therefore assume that the relaxed versions of \eqref{Tadpole_DpDpbar} that we will be writing in every specific case of our interest play an analogous role in guaranteeing UV-finiteness of string amplitudes. This is however not explicitly shown in our setup's.

If we now go back to purely supersymmetric brane configurations obtained by setting $N_{\overline{\text{D}p}}=0$ in equation \eqref{Tadpole_DpDpbar}, we find that it is actually possible to have both O$p^+$'s \& O$p^-$'s at the same time, as long as the constraint $ N_{\text{O}p^-}\geq N_{\text{O}p^+}$ is respected, in such a way that tadpole cancellation is realized without introducing susy breaking sources such as anti-branes. In this situation, tadpole cancellation would simply require adding the following amount of parallel D$p$-branes
\begin{equation}
N_{\text{D}p}\  \overset{!}{=}  \ 2^{p-4}\left( N_{\text{O}p^-} \, - \, N_{\text{O}p^+}\right)\ .
\end{equation}
In this most general setup, if we furthermore allow for the possibility that these objects be separated into smaller groups from one another, we find that the most general gauge group will be of the form
\begin{equation}
G_{\mathrm{YM}}\ = \ \left(\prod\limits_{a} \mathrm{U}(N_a)\right) \ \times \ \left(\prod\limits_{b} \mathrm{SO}(2N_b) \right)\ \times \ \left(\prod\limits_{c} \mathrm{USp}(2N_c) \right) \ .
\end{equation}
The corresponding total number of massless vector fields reads
\begin{equation}
\label{total_nv}
\nv \ \equiv \ \sum\limits_{a}N_a^2 \ + \ \sum\limits_{b}N_b(2N_b-1) \ + \ \sum\limits_{c}N_c(2N_c+1) \ ,
\end{equation}
which will precisely coincide with the bare quantity appearing in the lower dimensional supergravity description.
In the remainder of the paper, when discussing how these open string gaugings are embedded within the effective lower dimensional gauged supergravity theories, we will no longer specify an explicit form of the YM gauge group, nor specifically discuss concrete brane setup's. We hope to analyze concrete applications of the machinery presented here in the next future.

It is perhaps worth making one last general comment before moving to the technical supergravity analysis needed for our present purposes. It concerns the nature of the gauge groups just discussed here. Since in our analysis we will consider situations where the open and the closed string sectors are non-trivially coupled, the effective lower dimensional description will be a gauged supergravity with total gauge group featuring a mixing between $G_{\mathrm{YM}}$ \& $G_{\mathrm{ISO}}$, where $G_{\mathrm{YM}}$ is realized in terms of D-branes and O-planes as we have just seen, while $G_{\mathrm{ISO}}$ stems from the isometries of the bulk (internal) geometry. This suggests that the most general form of gauging within our 6D supergravity is expected to be a \emph{non-semisimple} extension of $G_{\mathrm{YM}}\times G_{\mathrm{ISO}}$\footnote{The extra non-vanishing mixed structure constants are roughly speaking due to the possibility of turning on RR \& NS-NS background fluxes, as well as YM flux associated with open string vector fields.}.

\section{Gauged $\mathcal{N}=(1,1)$ Supergravities in 6D}
\label{section:6D_sugra}
Ungauged $\mathcal{N}=(1,1)$ supergravity stems from dimensional reduction of type I supergravity on a $\mathbb{T}^{4}$. In this case, the complete set of closed string zero mode excitations is contained in the coupling between the gravity multiplet and four vector multiplets. Since the goal of this paper is that of using  $\mathcal{N}=(1,1)$ supergravities as a tool for studying type IIB orientifold reductions including an excited open string sector, we need to introduce their general formulation featuring the coupling with an arbitrary number of vector multiplets\footnote{We denote this number by $(4+\nv)$, where the first $4$ are needed in order to describe the closed string sector, while the extra $\nv$ accounts for the number of vector multiplets associated to the open string sector, just as appearing in equation \eqref{total_nv}.}. The (ungauged) theory enjoys the following global symmetry
\be
G_{\textrm{global}} \ = \ \mathbb{R}^{+}\,\times\,\mathrm{SO}(4,4+\nv) \ ,
\label{G_global}
\ee
where $\nv$ is the number of extra vector multiplets.
The physical (propagating) dof's of the theory are suitably rearranged into irrep's of the little group and of the global symmetry group as described in Table \ref{Table:dofs}.
\begin{table}[http!]
\renewcommand{\arraystretch}{1}
\begin{center}
\scalebox{1}[1]{
\begin{tabular}{|c|c|c|c|}
\hline
6D fields & $\mathrm{SO}(4)=\mathrm{SU}(2)_{\textrm{L}}\times \mathrm{SU}(2)_{\textrm{R}}$ & $\mathbb{R}^{+}_{\Sigma}\,\times\,\mathrm{SO}(4,4+\nv)$ irrep's & \# dof's   \\
\hline \hline
$g_{\mu\nu}$ & $(\textbf{3},\textbf{3})$ & $\textbf{1}^{(0)}$ & $9$ \\
\hline
$\mathcal{B}_{\mu\nu}$ & $(\textbf{1},\textbf{3})\oplus (\textbf{3},\textbf{1})$ & $\textbf{1}^{(+2)}$ & $6$ \\
\hline
$\mathcal{A}_{\mu}^{M} $ & $(\textbf{2},\textbf{2})$ & ${\tiny\yng(1)}^{(-1)}\equiv (\textbf{8}+{\nv})^{(-1)}$ & $4(8+\nv)$ \\
\hline 
$\Sigma$ & $(\textbf{1},\textbf{1})$ & $\textbf{1}^{(-1)}$ & $1$ \\
\hline 
$\mathcal{V}_{M}{}^{\underline{M}}$ & $(\textbf{1},\textbf{1})$ & $\textbf{adj}^{(0)}$ & $4(4+\nv)$ \\
\hline
\end{tabular}
}
\end{center}
\caption{\it The $(64+8\nv)_{B}$ bosonic dof's of the theory arranged into irrep's of $\mathrm{SO}(4)_{\textrm{little}}\times G_{\textrm{global}}$, the internal global symmetry being the one defined in  \protect\eqref{G_global}. Note that, within the scalars transforming in the adjoint, one should subtract the compact generators to get the correct number of propagating dof's.} \label{Table:dofs}
\end{table}

In particular, the $17+4\nv$ scalar fields of the theory parametrize the following coset geometry
\be
\mathcal{M}_{\textrm{scalar}} \ = \ \underbrace{\mathbb{R}^{+}}_{\Sigma}\,\times\,\underbrace{\frac{\mathrm{SO}(4,4+\nv)}{\mathrm{SO}(4)\times\mathrm{SO}(4+\nv)}}_{\mathcal{H}_{MN}} \ ,
\label{G_global_coset}
\ee
where the scalar coset representative $\mathcal{H}_{MN}$ is written in terms of a vielbein $\mathcal{V}_{M}{}^{\underline{M}}$ as 
\be
\mathcal{V}_{M}{}^{\underline{M}}\mathcal{V}_{N}{}^{\underline{M}}\,\equiv\,\mathcal{V}_{M}{}^{\underline{m}}\mathcal{V}_{N}{}^{\underline{m}}\,+\,\mathcal{V}_{M}{}^{\hat{\underline{m}}}\mathcal{V}_{N}{}^{\hat{\underline{m}}}\,+\,\mathcal{V}_{M}{}^{\underline{I}}\mathcal{V}_{N}{}^{\underline{I}}\,=\,\mathcal{H}_{MN}\ ,
\ee
where the local $\mathrm{SO}(4)\times\mathrm{SO}(4+\nv)$ index $\underline{M}$ has been split into $(\underline{m},\hat{\underline{m}},\underline{I})$, related to its $\mathrm{SO}(4)_{\textrm{timelike}}$, $\mathrm{SO}(4)_{\textrm{spacelike}}$, and $\mathrm{SO}(\nv)$ parts, respectively. The kinetic Lagrangian is given by
\be
\mathcal{L}_{\textrm{kin}}\,=\,-2\Sigma^{-2}(\partial\Sigma)^2
\,+\, \frac{1}{16}\,\partial\mathcal{H}_{MN}\partial\mathcal{H}^{MN} \ . 
\ee

The deformations of the ungauged theory which are consistent with bosonic symmetry as well as supersymmetry can arranged into the following embedding tensor irrep's
\be
\Yvcentermath1
\Theta \ \in \ \underbrace{{\tiny\yng(1)}^{(+3)}}_{\zeta_{M}} \, \oplus \,\underbrace{{\tiny\yng(1)}^{(-1)}}_{\xi_{M}} \, \oplus \,\underbrace{{\tiny\yng(1,1,1)}^{(-1)}}_{f_{[MNP]}} \ , 
\label{Theta_irreps}
\ee
where $\zeta_{M}$ corresponds to a massive deformation inducing a St\"uckelberg coupling for the two-form $\mathcal{B}_{\mu\nu}$, while the remaining two irreducible pieces are traditional gaugings. In particular, $f_{MNP}$ purely gauges a subgroup of $\mathrm{SO}(4,4+\nv)$, whereas $\xi_{M}$ gauges a combination of the $\mathbb{R}^+_{\Sigma}$ generator and generators in the $\mathrm{SO}(4,4+\nv)$ part.

Now, given a specification of the embedding tensor $\Theta$ transforming as in \eqref{Theta_irreps}, the consistency of the deformed theory demands its gauge invariance, which is enforced by imposing the following set of quadratic constraints (QC)
\be
\begin{array}{rcccccccrccc}
3\,f_{R[MN}\,f_{PQ]}{}^{R}\,-\,2\,f_{[MNP}\,\xi_{Q]} & = & 0 & , & & & & & \zeta_{(M}\,\xi_{N)} & = & 0 & , \\[2mm]
f_{MNP}\,\zeta^{P} \,-\, \xi_{[M}\,\zeta_{N]} & = & 0 & , & & & & & \xi_{M}\,\xi^{M} & = & 0 & , \\[2mm]
f_{MNP}\,\xi^{P} & = & 0 & , & & & & & \zeta_{M}\,\xi^{M} & = & 0 & ,
\end{array}
\label{Full_QC}
\ee
which include conditions for the closure of the gauge algebra, \emph{i.e.} generalized Jacobi identities. In \eqref{Full_QC}, contractions are defined by means of the invariant $\mathrm{SO}(4,4+\nv)$ metric $\eta_{MN}$ and its inverse $\eta^{MN}$. In what follows, we will perform a lightcone (LC) basis choice within the $\mathrm{SO}(4,4)$ sector, combined with a standard Cartesian basis along the remaining $\mathrm{SO}(\nv)$ directions. The explicit form of $\eta$ in this case is
\be
\eta_{MN} \ = \ \left(
\begin{array}{c|c|c}
\mathbb{O}_{4} &\mathbb{I}_{4} &\mathbb{O}_{4,\nv} \\ \hline
\mathbb{I}_{4} &\mathbb{O}_{4} &\mathbb{O}_{4,\nv} \\ \hline
\mathbb{O}_{\nv,4} &\mathbb{O}_{\nv,4} &\mathbb{I}_{\nv}
\end{array}
\right) \ .
\ee
It is perhaps worth mentioning that this choice of basis precisely matches the one made in \cite{Dibitetto:2019odu} within the $(4,4)$ part, which will represent the closed string sector of our type IIB orientifold compactifications. This choice is justified by the fact that closed string background fluxes have a natural mapping into LC components of the embedding tensor.

Embedding tensor deformations turn out to induce Yukawa-like couplings between scalars and fermions, which are parametrized by the so-called fermionic shift matrices. As a consequence, supersymmetry invariance of the action requires the presence of a scalar potential, which turns out to be quadratic in $\Theta$. Its explicit form in terms of embedding tensor irrep's reads
\begin{align}
V \,=\, \frac{g^2}{4} \, & \bigg[ f_{MNP}\,f_{QRS}\Sigma^{-2}  \left(\frac{1}{12}\mathcal{H}^{MQ}\mathcal{H}^{NR}\mathcal{H}^{PS}-\frac{1}{4}\mathcal{H}^{MQ}\eta^{NR}\eta^{PS}+\frac{1}{6}\eta^{MQ}\eta^{NR}\eta^{PS}\right)\notag\\[1mm] 
& +\frac{1}{2}\zeta_{M}\zeta_{N}\Sigma^{6}\mathcal{H}^{MN}\,+\,\frac{2}{3} f_{MNP}\zeta_{Q}\Sigma^{2}\mathcal{H}^{MNPQ}\,+\,\frac{5}{4}\xi_{M}\xi_{N}\Sigma^{-2}\mathcal{H}^{MN} \bigg] \ ,
\label{Full_VSugra}
\end{align}
where $\mathcal{H}^{MN}$ denotes the inverse of $\mathcal{H}_{MN}$ and $g$ is the gauge coupling constant. For simplicity, in the remainder of the paper, we fix $g=2$. The four-index antisymmetric object $\mathcal{H}^{MNPQ}$ appearing above is instead defined through
\be
\mathcal{H}_{MNPQ} \, \equiv \, \epsilon_{\underline{m}\underline{n}\underline{p}\underline{q}}\,\mathring{\mathcal{V}}_{M}{}^{\underline{m}}\mathring{\mathcal{V}}_{N}{}^{\underline{n}}\mathring{\mathcal{V}}_{P}{}^{\underline{p}}\mathring{\mathcal{V}}_{Q}{}^{\underline{q}} \ ,
\ee
in terms of the Cartesian vielbein $\mathring{\mathcal{V}}_{M}{}^{\underline{M}}$, which is in turn related to the LC one $\mathcal{V}_{M}{}^{\underline{M}}$ through $\mathring{\mathcal{V}}_{M}{}^{\underline{M}} \, = \, \mathcal{V}_{M}{}^{\underline{N}}\,U_{\underline{N}}{}^{\underline{M}}$, with
\be
U_{\underline{M}}{}^{\underline{N}} \ = \ \left(
\begin{array}{c|c|c}
-\frac{1}{\sqrt{2}}\mathbb{I}_{4} &\frac{1}{\sqrt{2}}\mathbb{I}_{4} &\mathbb{O}_{4,\nv} \\ \hline
\frac{1}{\sqrt{2}}\mathbb{I}_{4} &\frac{1}{\sqrt{2}}\mathbb{I}_{4} &\mathbb{O}_{4,\nv} \\ \hline
\mathbb{O}_{\nv,4} &\mathbb{O}_{\nv,4} &\mathbb{I}_{\nv}
\end{array}
\right) \ ,
\ee
transforming the LC metric into $\mathrm{diag}(-\mathbb{I}_{4},+\mathbb{I}_{4},\mathbb{I}_{\nv})$.

In \cite{Dibitetto:2019odu} all possible orientifold reductions yielding $\mathcal{N}=(1,1)$ theories in six dimensions were studided within the closed string sector. The closed string dynamics turned out to be contained within the theories with only the four (universal) vector multiplets included. In the next sections we will select the type IIB cases of interest and include an excited open string sector. This will require analyzing the 6D supergravity theories in the form presented in this section, \emph{i.e.} with the inclusion of $\nv$ extra vector multiplets. Such an extension will allow us to study open string dof's like brane position moduli and/or Wilson line moduli, \emph{i.e.} axions arising from internal legs of the worldvolume gauge fields. Moreover, we will be able to consider possible physical effects of a non-Abelian worldvolume theory, and/or the presence of worldvolume flux wrapping internal space. 

\section{O$5$/D$5$ \& Open Strings}
\label{section:O5}

\noindent Let us start by considering the minimal possible spacetime filling O-planes that respect 6D Lorentz symmetry, \emph{i.e.} O$5$-planes. These are placed as follows within 10D spacetime,
\be
\mathrm{O}5 \ : \quad
\underbrace{\times \ \times \ \times \ \times \ \times \ \times}_{\textrm{6D spacetime}} \ | \ \underbrace{- \ - \ - \ -}_{y^{m}} \ , \qquad 
\sigma_{\mathrm{O}5} \ : \ y^{m} \ \longrightarrow \ -y^{m} \ ,
\notag
\ee
where $\sigma_{\mathrm{O}5}$ is the orientifold involution, whose action flips the sign of all transverse coordinates. The O$5$ projection is realized at the level of the 10D supergravity fields by means of the simultaneous action of the aforementioned involution, together with the worldsheet parity operator. Such a procedure yields the correct field content of a half-maximal supergravity in 6D. The resulting details of this projection are collected in Table \ref{Table:O5_Projection}.
\begin{table}[http!]
\renewcommand{\arraystretch}{1}
\begin{center}
\scalebox{1}[1]{
\begin{tabular}{|c|c|c|c|}
\hline
IIB fields & $\sigma_{\mathrm{O}5}$ & $\Omega$ & \# dof's   \\
\hline \hline
$e^{m}{}_{n}$ & $+$ & $+$ & $16-6\,=\,10$ \\
\hline 
$B_{mn}$ & $+$ & $-$ & --- \\
\hline
$\Phi$ & $+$ & $+$ & $1$ \\
\hline \hline
$C_{(0)}$ & $+$ & $-$ & --- \\
\hline 
$C_{mn}$ & $+$ & $+$ & $6$ \\
\hline 
$C_{mnpq}$ & $+$ & $-$ & --- \\
\hline \hline
$Y^{Im}$ & $-$ & $-$ & $4\nv$ \\
\hline
\end{tabular}
}
\end{center}
\caption{\it The $\mathbb{Z}_{2}$ parity of all internal components of the different IIB fields in the presence of spacetime filling O$5$-planes. The allowed ones yield excitable 6D scalar fields. Note that the total amount of resulting scalars correctly gives $17+4\nv$, \emph{i.e.} the dimension of the supergravity coset \protect\eqref{G_global_coset}.} \label{Table:O5_Projection}
\end{table}

In this case, due to presence of O$5$'s and D$5$'s, the reduction Ansatz can be formulated in a $\mathrm{SL}(4,\mathbb{R})\times G_{\mathrm{YM}}$ covariant way. The 10D bulk supergravity Ansatz containing the $17$ closed string scalars reads
\begin{align}
ds_{(10)}^2=&\ \tau^{-2}\,g_{\mu\nu}dx^{\mu}dx^{\nu} \, +\, \rho \, M_{mn} \, dy^{m}dy^{n} \ , \\[1mm]
e^{\Phi}=&\ \rho\,\tau^{-2}\ , \\[1mm]
B_{(2)} = &\ \frac{1}{6} \, \epsilon_{mnpq}\,h^{m}y^{n}dy^{p}\wedge dy^{q} \, + \, \dots\ , \\[1mm]
\label{eq:O5C0}
C_{(0)} = &\ f_{m}\,y^{m} \, + \, \dots\ , \\[1mm]
C_{(2)} = &\ \frac{1}{(2!)^2} \, \epsilon_{mnpq}\,\gamma^{mn}dy^{p}\wedge dy^{q}\, + \, \dots \ , \\[1mm]
C_{(4)} = &\ 0\, + \, \dots \ ,
\end{align}
where ``$\, + \, \dots$'' denotes that we are discarding the terms in the Ansatz that do not contribute to the scalar potential. The scalars $\rho$, $\tau$ represent the volume and dilaton would-be moduli, $M_{mn}$ is an element of $\mathrm{SL}(4,\mathbb{R})/\mathrm{SO}(4)$ describing deformations of the internal metric, and $\gamma^{mn}$ is antisymmetric and contains the scalars coming from the R-R two-form $C_{(2)}$. These modes add up to $17$, as they should. The remaining $4\nv$ scalars are part of the open-string sector and are denoted as $Y^{Im}$, where the index $I$ labels the adjoint representation of $G_{\rm{YM}}$.

On the other hand, $h^{m}$ and $f_{m}$ are constants parametrizing the $\overline H_{(3)}$ and $\overline F_{(1)}$ fluxes within the closed string sector,
\be
\overline{F}_{m} \ = \ f_{m} \ , \qquad \overline{H}_{mnp} \ = \ \epsilon_{mnpq}\,h^{q} \ ,
\ee
while in the open-string one we have the possibility of considering a non-Abelian gauge group with structure constants $g_{IJ}{}^{K}$. The consistency requirements on the aforementioned flux parameters purely reduce to the Jacobi identity for the Yang-Mills structure constants,
\be
g_{[IJ}{}^{I'}\,g_{K]I'}{}^{L} \ \overset{!}{=} \ 0 \ .
\label{Jacobi_gIJK}
\ee
Note that the flux tadpole induced by $H_{(3)}$ and $F_{(1)}$ does not have to vanish, since the Bianchi identity for $C_{(2)}$ gets modified by the presence of O$5$/D$5$ sources:
\be
\underbrace{dF_{(3)}}_{=\,0} \, - \, \underbrace{H_{(3)}\wedge F_{(1)}}_{\neq\,0} \ = \ j_{(4)}^{\textrm{O5/D5}} \ ,
\ee
where $ j_{(4)}^{\textrm{O5/D5}} = Q_{5}\,\mathrm{vol}_{\mathcal{M}_4}$ is the effective current density. In the case at hands, we have that 
\be
Q_{5}=2\kappa^2\left(N_{\Dp5}\,\mu_{\textrm{D}5}+\mu_{\textrm{O}5}\right)
=2\kappa^2 \mu_{\textrm{D}5}\left(N_{\Dp5}+2\epsilon_{{\rm O}5}\right)
\,.
\ee
In the last equality we have made use of the relation between the D5 and O5 charges given in \eqref{eq:Optension/charge}. The sign $\epsilon_{\rm O5}=\pm 1$ precisely determines the type of O$5^{\pm}$ plane that we are considering.\footnote{We refer to eq.~\eqref{eq:epsilonOp} for further details.} The tadpole condition imposes the following condition
\be \label{eq:D5O5tadpole}
 h^{m}\, f_{m} \ \overset{!}{=} \, Q_{5}\,=\,2\kappa^2 T_{\rm D5}\left(N_{\Dp5}+2\epsilon_{{\rm O}5}\right)\, ,
\ee
which must be taken into account when matching the scalar potential of the compactification with that of supergravity, as the latter only knows about the fluxes.

The scalar potential of the compactification ignoring the open-string sector was previously computed in \cite{Dibitetto:2019odu}. Now we build on their results and also take into account open-string effects. The worldvolume action of the D5 branes contains two pieces: the DBI and the WZ actions. The first directly gives a contribution to the scalar potential, while the contribution of the second secretly appears through the bulk scalar potential given in Appendix~\ref{app:bulkreduction}. The reason lies in the fact that the WZ action contains couplings between the open-string fields and the R-R potentials which in turn give rise to modified (bulk) field strengths associated to the dual R-R potentials.

Let us consider each contribution separately, first focusing on the one coming from the DBI. The DBI action of the D5 branes is given by (see Appendix~\ref{app:iib-DBI} for a detailed description)
\be
S_{\textrm{D}5}^{\textrm{DBI}} \, = \, -T_{\textrm{D}5}\int{d^{6}x\, \mathrm{Tr}\left(e^{-\hat{\Phi}}\sqrt{-\mathrm{det}(\mathbb{M}_{MN})\,\mathrm{det}(\mathbb{Q}^{i}{}_{j})}\right)} \ ,
\label{eq:DBI-D5brane}
\ee
where $T_{\textrm{D}5}=2\pi \ell_s^{-6}$ is the D5 brane tension. The indices $M, N, \dots$ are worldvolume indices whereas $i, j, \dots$ denote the transverse ones. The matrices $\mathbb{M}$ and $\mathbb{Q}$ are defined as
\begin{align}
\mathbb{M}_{MN}
=&\
\ \mathrm{P}\left[
	\hat E_{MN}
	+\hat E_{Mi}(\mathbb{Q}^{-1}-\delta)^{ij}\hat E_{jN}
	\right]
	+\lambda \mathcal{F}_{MN}
\ ,
\\
\mathbb{Q}^i{}_j
=&\
\delta^i{}_j +i \lambda [Y^i,Y^k] \hat E_{kj}
\ ,
\end{align}
where ${E}_{{\cal M}{\cal N}}={\hat G}_{{\cal M}{\cal N}}+{\hat B}_{{\cal M}{\cal N}}$.\footnote{${\cal M}, {\cal N}, \dots$ are ten-dimensional indices and the meaning of the hat on ten-dimensional fields is explained in \eqref{eq:Taylor}.}
Making use of the above compactification Ansatz, one finds that the matrices $\mathbb{M}$ and $\mathbb{Q}$ are given by
\begin{align}
\label{eq:D5Mmunu}
\mathbb{M}_{\mu\nu} = & \ \tau^{-2}\,g_{\mu\nu} \,+\, \dots \ ,\\[1mm]
\label{eq:D5Qmn}
\mathbb{Q}^{m}{}_{n} = & \ \delta^{m}_{n} - \lambda\,\rho\, g_{IJ}{}^{K}Y^{Im}Y^{Jp}M_{pn}t_{K}+\frac{\lambda^2}{3}\, \epsilon_{npqr} h^{r}\,g_{IJ}{}^{K}Y^{Im}Y^{Jp}Y^{Lq}t_{K}t_{L}
\,+\,\dots \ ,
\end{align}
where now the dots mean that we are ignoring terms that do not contribute to the scalar potential and also the ones which are of higher-order in $\lambda$. The generators of $G_{\rm YM}$ are denoted by $t_I$, and our conventions are such that $[t_I,t_J]=i\, g_{IJ}{}^K \, t_K$. Making use of \eqref{eq:D5Mmunu} and \eqref{eq:D5Qmn} in \eqref{eq:DBI-D5brane}, we obtain the following contribution to the scalar potential,
\begin{equation}\label{eq:VDBI-D5}
\begin{aligned}
V^{\rm{DBI}}_{\rm{D}5}=\,&\rho^{-1}\tau^{-4}\left[2\kappa_6^2N_{\Dp5}T_{\Dp 5}+\frac{2\kappa_6^2\lambda^2N_{\Dp5}T_{\rm D5}}{6}g_{IJK}\epsilon_{mnpq}h^{q}Y^{Im}Y^{Jn}Y^{Kp}\right]\\[1mm]
&+\rho\tau^{-4}\left(\frac{2\kappa_6^2\lambda^2N_{\Dp5}T_{\rm D5}}{4} g_{IJ}{}^{M}g_{KLM}M_{mn}M_{pq}Y^{Im}Y^{Jp}Y^{Kn}Y^{Lq}\right)\,,
\end{aligned}
\end{equation}
where $2\kappa^2_{6}=16\pi G_6$, being $G_6$ the six-dimensional Newton's constant. 

In addition to \eqref{eq:VDBI-D5}, we will have the analogous contribution from the O5. Since the latter is non-dynamical, its contribution merely reduces to tension term (namely, the first one in \eqref{eq:VDBI-D5}). Hence, the total DBI contribution from both types of sources is 
\begin{equation}\label{eq:VDBI-D5+O5}
\begin{aligned}
V^{\rm{DBI}}_{\Op5/\Dp5}=\,&\rho^{-1}\tau^{-4}\left[2\kappa_6^2\left(N_{\Dp5}T_{\Dp 5}+T_{\Op5}\right)+\frac{2\kappa_6^2\lambda^2N_{\Dp5}T_{\rm D5}}{6}g_{IJK}\epsilon_{mnpq}h^{q}Y^{Im}Y^{Jn}Y^{Kp}\right]\\[1mm]
&+\rho\tau^{-4}\left(\frac{2\kappa_6^2\lambda^2N_{\Dp5}T_{\rm D5}}{4} g_{IJ}{}^{M}g_{KLM}M_{mn}M_{pq}Y^{Im}Y^{Jp}Y^{Kn}Y^{Lq}\right)\,,
\end{aligned}
\end{equation}

On the other hand, the Wess-Zumino action contains a coupling between the scalars $Y^{Ii}$ and $C_{(8)}$, see \eqref{eq:couplingtoC8}. As shown in Appendix~\ref{app:O5D5}, this can be understood through a modified field strength $F_{m}$ of the form,
\begin{equation}\label{eq:D5O5Fmodified}
F_{m} = \, \overline{F}_{m} -  \frac{\tilde\lambda_5^2}{3!}\,\epsilon_{mnpq}\, g_{IJK} \,Y^{In}\,Y^{Jp}\,Y^{Kq} \, ,
\end{equation}
where $\tilde\lambda_5=(2\kappa_6^2 N_{\rm D5} T_{{\rm D5}})^{1/2} \lambda$ and $g_{IJK}=g_{IJ}{}^L\, \kappa_{LK}$. Crucially now $F_{m}$ is not pure flux, as it has the second contribution from the open-string scalars. On the contrary, the field strength $H_{mnp}$ is not modified, so it simply reads
\begin{equation}
H_{mnp} =\, \overline{H}_{mnp} \, .
\end{equation}

Finally, all that is left is to take the expression for the bulk scalar potential computed in \cite{Dibitetto:2019odu} (see also Appendix~\ref{app:bulkreduction}) and replace, according to \eqref{eq:D5O5Fmodified},
\begin{equation}
{\overline F}_{m}\to\overline{F}_{m} -  \frac{\tilde\lambda_5^2}{3!}\,\epsilon_{mnpq}\, g_{IJK} \,Y^{In}\,Y^{Jp}\,Y^{Kq} \, .
\end{equation}
The resulting expression has to be added to \eqref{eq:VDBI-D5+O5}, which yields the following expression for the full scalar potential, 
\begin{align}
V_{\textrm{O}5/\textrm{D}5} 
= & \ 
\frac{1}{2}\, \rho^{-3}\tau^{-2}\, H_{mnp}\, H_{m'n'p'} \, M^{mm'}\, M^{nn'}\, M^{pp'}\,
+\,\frac12 \, \rho\,\tau^{-6}\, F_{m}\, F_{m'}\, M^{mm'}
\nonumber
\\[5pt]
& 
+\,  \frac{{\tilde \lambda}^2_5}{12}\,\tau^{-4}\,g_{IJK}\,\left[
	2\, \rho^{-1} H_{mnp}  Y^{Im}Y^{Jn}Y^{Kp}
 \,+\, 3\,\rho\, g_{I'J'}{}^{K}Y^{Im}Y^{Jn}Y^{I'p}Y^{J'q}M_{mn}M_{pq}	
	\right] 
\nonumber\\[5pt]
&
+\,2\, \kappa_6^2\,\rho^{-1}\tau^{-4}	\, T_{\rm D5}\left(N_{\rm D5}+2\epsilon_{{\rm O}5}\right)\,	
	\ ,
\label{VIIB_O5}
\end{align}
where the bulk and the WZ contributions correspond to the first line, while the rest come from DBI.

\subsection*{Matching with $\mathcal{N}=(1,1)$ gauged supergravity}

Since the O$5$ reduction Ansatz respects $\mathrm{SL}(4,\mathbb{R})\times G_{\mathrm{YM}}$ covariance, the fundamental index $M$ of $\mathrm{SO}(4,4+\nv)$ is split accordingly
\be
\begin{array}{lclclclc}
M & \longrightarrow & m & \oplus & \bar{m} & \oplus & I & ,
\end{array}\notag
\ee
where $m$ and $\bar{m}$ are (anti)fundamental indices of $\mathrm{SL}(4,\mathbb{R})$, while $I$ is an adjoint index of $G_{\mathrm{YM}}$.
The dictionary between flux parameters and embedding tensor components is summarized in Table~\ref{Table:O5_Fluxes/ET}.
\begin{table}[http!]
\renewcommand{\arraystretch}{1}
\begin{center}
\scalebox{1}[1]{
\begin{tabular}{|c|c|c|}
\hline
IIB fluxes & $\Theta$ components &  Dictionary \\
\hline \hline
$\overline{H}_{mnp}=\,\epsilon_{mnpq}\,h^{q}$ & $\zeta_{\bar{m}}$ & $\zeta_{\bar{m}}=h^{m}$ \\
\hline
$\overline{F}_{m}=\,f_{m}$ & $f_{\bar{m}\bar{n}\bar{p}}$ & $ f_{\bar{m}\bar{n}\bar{p}}= \epsilon^{mnpq}\,f_{q}$ \\
\hline
$g_{IJ}{}^{K}$ & $f_{IJK}$ & $ f_{IJK}=\,{\tilde \lambda_5}^{-1}\,g_{IJ}{}^{L}\kappa_{KL}$ \\
\hline
\end{tabular}
}
\end{center}
\caption{\it The embedding tensor/fluxes dictionary for type IIB reductions with spacetime filling O5-planes and D5-branes. } \label{Table:O5_Fluxes/ET}
\end{table}
With these non-vanishing components of the embedding tensor, the general form of the QC \eqref{Full_QC} reduces to the Jacobi identities for the structure constants $g_{IJK}$, as given in \eqref{Jacobi_gIJK}.

In order to evaluate the general supergravity scalar potential \eqref{Full_VSugra} in our specific case, we need to identify how the $17+4\nv$ scalars parametrize $\Sigma$ and $\mathcal{H}_{MN}$ appearing there. This is done by directly expressing $\Sigma$ and the coset representative $\mathcal{V}_{M}{}^{\underline{M}}$ as a function of the scalars $\left(\rho,\tau,M_{mn},\gamma^{mn},Y^{Im}\right)$.
In particular, we find that
\be
\Sigma \, = \, \rho^{-1/2} \ ,
\ee
and the $\rm SO(4,4+\nv)$ coset element is
\be
\mathcal{V}_{M}{}^{\underline{M}} \, = \, \left(
\begin{array}{c|c|c}
\tau^{-1}L_{m}{}^{\underline{m}} & \mathbb{O}_{4} &  \mathbb{O}_{4,N} \\[2pt] \hline
\tau^{-1}\mathcal{C}^{mn}L_{n}{}^{\underline{m}} & \tau\,L^{m}{}_{\underline{m}} & -{\tilde \lambda_5}Y^{Im}\delta^{\underline{I}}_{I} \\[2pt] \hline
\tau^{-1}{\tilde \lambda_5}\delta_{IJ}Y^{Jm}L_{m}{}^{\underline{m}} &  \mathbb{O}_{N,4} & \delta^{\underline{I}}_{I} 
\end{array}
\right) \ ,
\ee
where $L_{m}{}^{\underline{m}}$ is an $\mathrm{SL}(4,\mathbb{R})/\mathrm{SO}(4)$ coset element satisfying $L_{m}{}^{\underline{m}}L_{n}{}^{\underline{m}}\overset{!}{=}M_{mn}$, while
\be
\mathcal{C}^{mn} \, \equiv \, \gamma^{mn} \, - \, \frac{{\tilde \lambda_5}^2}{2}Y^{Im}Y^{Jn}\delta_{IJ} \ .
\ee
By plugging the above parametrization of the scalars and embedding tensor into the general form of the scalar potential \eqref{Full_VSugra}, we find a perfect agreement with \eqref{VIIB_O5}, which was calculated from direct dimensional reduction.

\section{O$7$/D$7$ \& Open Strings}
\label{section:O7}

\noindent Let us now consider reductions of type IIB with the inclusion of spacetime filling O$7$-planes. These are placed as follows within ten-dimensional spacetime
\be
\mathrm{O}7 \ : \quad
\underbrace{\times \ \times \ \times \ \times \ \times \ \times}_{\textrm{6D spacetime}} \ | \ \underbrace{\underbrace{\times \ \times}_{y^{a}} \, \underbrace{- \ -}_{y^{i}}}_{y^m} \ , \qquad 
\sigma_{\mathrm{O}7} \ : \ y^{i} \ \longrightarrow \ -y^{i} \ ,
\notag
\ee
where $\sigma_{\mathrm{O}7}$ is the orientifold involution, whose action flips the sign of all transverse coordinates, while leaving the $y^{a}$ internal coordinates invariant. The O$7$ projection is realized at the level of the ten-dimensional supergravity fields by means of the simultaneous action of the aforementioned involution, together with the fermionic number $(-1)^{F_L}$ and the worldsheet parity operator. Such a procedure yields the correct field content of a half-maximal supergravity in six dimensions. The resulting details of this are collected in Table \ref{Table:O7_Projection}.
\begin{table}[http!]
\renewcommand{\arraystretch}{1}
\begin{center}
\scalebox{1}[1]{
\begin{tabular}{|c|c|c|c|}
\hline
IIB fields & $\sigma_{\mathrm{O}7}$ & $(-1)^{F_L}\Omega$ & \# dof's   \\
\hline \hline
$e^{a}{}_{b}\,\oplus\,e^{i}{}_{j}$ & $+$ & $+$ & $2(4-1)\,=\,6$ \\
\hline 
$e^{a}{}_{j}\,\oplus\,e^{i}{}_{b}$ & $-$ & $+$ & --- \\
\hline
$B_{ai}$ & $-$ & $-$ & $4$ \\
\hline
$B_{ab}$ & $+$ & $-$ & --- \\
\hline
$B_{ij}$ & $+$ & $-$ & --- \\
\hline
$\Phi$ & $+$ & $+$ & $1$ \\
\hline \hline
$C_{(0)}$ & $+$ & $+$ & 1 \\
\hline 
$C_{ai}$ & $-$ & $-$ & $4$ \\
\hline 
$C_{ab}$ & $+$ & $-$ & --- \\
\hline
$C_{ij}$ & $+$ & $-$ & --- \\
\hline
$C_{abij}$ & $+$ & $+$ & $1$ \\
\hline \hline
$Y^{Ii}$ & $-$ & $-$ & $2\nv$ \\
\hline
$\mathcal{A}^{I}{}_{a}$ & $+$ & $+$ & $2\nv$ \\
\hline
\end{tabular}
}
\end{center}
\caption{\it The $\mathbb{Z}_{2}$ parity of all internal components of the different IIB fields in the presence of spacetime filling O$7$-planes. The allowed ones yield excitable 6D scalar fields. Note that the total amount of resulting scalars correctly gives $17+4\nv$, \emph{i.e.} the dimension of the supergravity coset \protect\eqref{G_global_coset}.} \label{Table:O7_Projection}
\end{table}

Now, because the embedding of O$7$-planes breaks internal diffeomorphism covariance, the reduction scheme can only be formulated in a $\mathrm{SL}(2,\mathbb{R})_{a}\times\mathrm{SL}(2,\mathbb{R})_{i}\times G_{\mathrm{YM}}$ covariant way. The ten-dimensional supergravity Ansatz containing the $17$ closed string scalars reads
\begin{align}
d{s}^2=&\ \tau^{-2}\,g_{\mu\nu}dx^{\mu}dx^{\nu} \, +\, \rho \, \left(\sigma^{2}M_{ab} \, v^{a}v^{b}\,+\,\sigma^{-2}M_{ij} \, v^{i}v^{j} \right)\ , \\[1mm]
e^{\Phi}=&\ \rho\,\tau^{-2}\ , \\[1mm]
{B}_{(2)} =&\ B_{ai}\, v^{a}\wedge v^{i} \,+ \, {\beta} \, + \, \dots\ , \\[1mm]
\label{eq:C0}
{C}_{(0)} =&\ \chi \,+\, \alpha\, + \, \dots \ , \\[1mm]
\label{eq:C2}
{C}_{(2)} =&\ C_{ai}\, v^{a}\wedge v^{i}  \,+ \, C_{(0)} \,B_{ai}\, v^{a}\wedge v^{i}\, +\, \gamma \, + \, \dots\ , \\[1mm]
{C}_{(4)} =&\ \frac{\psi}{(2!)^2} \, \epsilon_{ab}\epsilon_{ij}\,v^{a}\wedge v^{b} \wedge v^{i}\wedge v^{j}  \, + \, \dots \ ,
\end{align}
where again ``$\, + \, \dots$'' means that we are discarding the terms in the Ansatz that do not contribute to the scalar potential. The scalars $\rho$ and $\tau$ still represent the volume and dilaton would-be moduli, $\sigma$ is a non-universal geometric deformation controlling the relative size between the two-cycle wrapped by the O$7$ and the transverse one, $M_{ab}$ and $M_{ij}$  are elements of $\mathrm{SL}(2,\mathbb{R})_{a(i)}/\mathrm{SO}(2)$, and, finally, $B_{ai}$, $C_{ai}$, $\chi$, $C_{abij}=\epsilon_{ab} \epsilon_{ij} \psi$ are axionic scalars coming from the NS-NS two-form and the R-R forms. These modes again add up to $17$, as expected. Since in this case the metric flux is allowed by the O$7$ involution, we have introduced a parallelization of the internal manifold with torsion given in terms of the Maurer-Cartan one-forms $v^{m}=v^{m}{}_{\underline{m}} \, dy^{\underline{m}}$. These satisfy
\be
dv^{m}\,+\,\frac{1}{2}\,\omega_{np}{}^{m}v^{n}\wedge v^{p} \, = \, 0 \ ,
\ee
for some constants $\omega_{np}{}^{m}$, which turn out to be the structure constants of the underlying Lie algebra and therefore fulfill the Jacobi identities as an integrability condition,
\begin{align}
\omega_{[mn}{}^r\, \omega_{p]r}{}^q
= 0 \ .
\label{eq:Jacobi-omega}
\end{align}

The 0-form $\alpha=\alpha(y)$ entering in the reduction Ansatz of $ C_{(0)}$, given in \eqref{eq:C0}, introduces the 1-form flux $\overline{F}_a$ via its exterior derivative:
\begin{align}
d\alpha=\overline{F}_a\, v^a\ ,
\end{align}
with $\overline{F}_{a}=f_a$. The integrability condition implies that
\begin{align}
f_a\, \eta_b\, \epsilon^{ab}
=
0
\ .
\label{eq:integrabilityF1}
\end{align}

The 2-form $\beta=\frac{1}{2}\beta_{ab}\, v^a\wedge v^b$ in the reduction Ansatz of ${B}_{(2)}$ does not give rise to further scalars but rather to fluxes, as it satisfies:
\begin{equation}
d\beta=\frac{1}{2}\, \overline{H}_{abi} \, v^a\wedge v^b\wedge v^i\, ,
\end{equation}
where $\overline{H}_{abi}=h_{i}\, \epsilon_{ab}$ denotes the $H$-flux. The integrability condition of this equation is trivially satisfied. 

Finally, the 2-form $\gamma= \gamma_{ai} \, v^a\wedge v^i$ in \eqref{eq:C2} encodes the 3-form flux $\overline F_{abi}$ via its exterior derivative,
\begin{align}
d\gamma=\frac{1}{2} \overline F_{abi} \, v^a\wedge v^b\wedge v^i \ ,
\end{align}
with $\overline{F}_{abi}=\epsilon_{ab}\, f_{i}$.

If we now add a stack of $N_{\Dp7}$ D$7$-branes parallel to the O$7$, these will require the existence of $\nv$ extra vector multiplets, labelled by the index $I$. In terms of 6D dof's, besides the new vector fields, our resulting gauged supergravity  will again have $4\nv$ new scalar modes. Half of them, denoted as $Y^{I{\underline i}}$, correspond to the scalar fields living in the $\nv$ vector supermultiplets, whereas the remaining half come from the internal components of the worldvolume gauge fields ${\cal A}^I{}_{a}$. This  way, the set $\left(Y^{Ii},\mathcal{A}^{I}{}_{a}\right)$ exactly parametrizes $4\nv$ independent extra scalar modes. The compactification Ansatze for the worldvolume gauge fields and the scalar fields are
\begin{align}
\mathcal{A}^{I} \, =\, \mathcal{A}^{I}{}_{a}{v}^{a}_{0}  \,+\, \sigma^{I}\, +\ldots \ ,
\qquad\qquad
Y^{I\underline{i}} \, =\, Y^{Ii} \, (v_0^{-1})^{\underline{i}}{}_i \ ,
\end{align}
where $v^{m}_{0}=v^{m}{}_{\underline m}|_{y^i=0}\,dy^{\underline m}$, with $m=(a,i)$, are the Maurer-Cartan 1-forms restricted to the worldvolume of the D7 branes\footnote{This is nothing but the leading-order contribution in $\lambda$ of the pull-back of $e^a$ onto the D7 worldvolume, which is all supergravity can capture.} and $\sigma^I$ is a 1-form satisfying 
\begin{equation}\label{eq:conditionsigmas}
g_{IJ}{}^{K}\, \sigma^{I}=0\, ,
\end{equation}
where $g_{IJ}{}^K$ are the Yang-Mills structure constants satisfying  the Jacobi identities:
\begin{align}
g_{[IJ}{}^{I'} \, g_{K]I'}{}^L = 0
\ .
\end{align}
In addition, $\sigma^I$ gives rise to the flux $\overline{\cF}^I{}_{ab}$ through
\begin{equation}\label{eq:diffeqsigmas}
d\sigma^I= \tfrac{1}{2}\overline{\cF}^I{}_{ab}\, v^{a}_{0}\wedge v^{b}_{0}\, ,
\end{equation}
where $\overline{\cF}^I{}_{ab}=g^I\, \epsilon_{ab}$.
The above condition \eqref{eq:conditionsigmas} on $\sigma^I$ has to be imposed in order to remove undesired dependence on the internal coordinates. Taking a exterior derivative, we can express this condition in terms of the fluxes $g^I$ as follows: 
\begin{equation}\label{eq:QGgI}
g_{IJ}{}^{K}\, g^{I}\, =\, 0\, ,
\end{equation}
which tells us  that the Killing-Cartan metric,

\begin{equation}
\kappa_{IJ}=g_{IK}{}^{L}g_{JL}{}^{K}\,,
\end{equation}
must be degenerate when the fluxes $g^I$ are turned on. Consequently, this implies that the Lie algebra cannot be semisimple. As we are going to see in what follows, the constraint \eqref{eq:QGgI} will arise in supergravity as one of the QC, \eqref{eq:QCD7O7}. Let us further remark that the integrability condition of \eqref{eq:diffeqsigmas} is automatically satisfied without imposing further constraints on the fluxes $g^{I}$. Such circumstance is very particular of this case. Taking these considerations into account and the compactification Ansatz, we find that the internal components of the field strength ${\cal F}^{I}$ are given by 
\begin{equation}
{\cal F}^{I}{}_{ab}
=
\overline{\cal F}^I{}_{a  b }
-g_{JK}{}^{I}{\cal A}^{J}{}_{a}{\cal A}^{K}{}_{ b}
+2 {\cal A}^{I}{}_{[a}\,\eta_{b]} \, .
\end{equation}

Due to the presence of the O7 plane, the $\mathbb{Z}_2$ truncation realized by the product $\sigma_{\text{O}7}\, (-1)^{F_L}\Omega$ projects out, some internal fluxes. Table \ref{Table:O7_Fluxes/ET} shows the exhaustive list of fluxes that are projected in by this truncation, where we observe that the fluxes $\omega_{ab}{}^i$, $\omega_{ij}{}^k$, $\omega_{ai}{}^j$ are forbidden. As a consequence, the Maurer-Cartan 1-forms $v^m$ are required to depend on the internal coordinates $y^m$ in a very particular way. This, together with the consistency of the Ansatz when studying the D7 brane effective action imposes the following functional dependence of the twist matrices $v^m{}_{\underline{m}}$:
\begin{align}
v^m{}_{\underline{m}}
=
\begin{pmatrix}
v^{a}{}_{\underline b}(y^c) & v^{a}{}_{\underline j}(y^k)
\\[2mm]
\mathbb{O}_2 & v^{i}{}_{\underline j}(y^c)
\end{pmatrix}
\ .
\label{eq:D7twist}
\end{align}

The presence of the D7/O7 sources modifies the Bianchi identity of $C_{(0)}$ as follows,
\begin{equation}
dF_{(1)}=j_{(4)}^{\textrm{O7/D7}} \,,
\end{equation}
where $j_{(4)}^{\textrm{O7/D7}}$ is the effective 7-brane current. The above equation involves both the metric flux $\omega_{ij}{}^{a}$ and the $F_{(1)}$ flux, $\overline{F}_{a}$. Upon integration over the transverse space, one obtains the following tadpole condition
\begin{align}\label{eq:tadpoleD7O7}
\frac12 \epsilon^{ij}\, \omega_{ij}{}^a \overline{F}_a\stackrel{!}{=}Q_7
\ ,
\end{align}
where $Q_7$ is the total charge, receiving contributions both from the D7 branes and the O7,
\begin{align}
Q_7=T_{\rm D7}(N_{\rm D7}+8\epsilon_{\rm O7})
\ ,
\end{align}
where $\epsilon_{\rm O7}=\pm1$ amounts to considering the presence of O$7^\pm$ planes.
As a consequence of \eqref{eq:tadpoleD7O7}, we will obtain a non-vanishing contribution in the scalar potential which is proportional to the  effective tension, as we are about to see.

The procedure one has to follow to compute the scalar potential is exactly the same as in D5/O5 case studied in the previous section, so we will skip most of the details here. In order to evaluate the contribution from the DBI, we just need to know the matrices $\mathbb{M}$ and $\mathbb{Q}$ in \eqref{eq:DBI-Dbrane}. In the case at hands, these read 
\begin{align}
\mathbb{M}_{\mu\nu} = & \ \tau^{-2}\,g_{\mu\nu} \,+\, \ldots \ ,
\\[2mm]
\mathbb{M}_{ab} 
= & \ 
\rho\, \sigma^2 M_{a b}
+\lambda \, \mathbb{M}^{(1)I}{}_{ab}\, t_I
+\lambda^2 \, \rho\, \mathbb{M}^{(2)IJ}{}_{ab}\, t_I \, t_J
+\ldots \ ,
\ ,
\nonumber\\[2mm]
\sqrt{\det\mathbb{Q}^{i}{}_{j}} 
= & \ 
1
+\frac{\lambda^2}{4}\rho^2\sigma^{-4} M_{ij} M_{i'j'} Y^{Ki} Y^{Li'} Y^{K'j} Y^{L'j'} g_{KL}{}^I  g_{K'L'}{}^{J}  t_{I} \, t_{J}
+\ldots \ ,
\end{align}
where
\begin{align}
\mathbb{M}^{(1)I}{}_{ab}
=&\
	2 B_{[a| i}  Y^{I j}  \omega_{|b] j}{}^{i}
	+2 A^{J}{}_{[a} B_{b] i} Y^{K i}  g_{JK}{}^{ I}
\nonumber\\[3pt]
&\
\qquad
	+   Y^{I i}\, \overline{H}_{abi}
	- B_{[a| i} B_{|b] j} 
	Y^{J i} Y^{K j} g_{JK}{}^I
	+ Y^{Ii}\, \eta_{[a}\, B_{b]i}
	+{\cal A}^{I}{}_{[a}\,\eta_{b]}
	+\cF^I{}_{ab}
\ ,
\label{eq:D7M1ab}
\\[3pt]
\mathbb{M}^{(2)IJ}{}_{ab}
=&\
\sigma^{2}\, M_{c (a}\, Y^{I i}Y^{K j} \left(
	A^{L}{}_{b)}\, \omega_{i j}\,^{c} g_{L K}{}^{J} 
	+\delta^J{}_K\, \omega_{b) j}\,^{k}\omega_{k i}\,^{c}
	\right)
\nonumber\\[3pt]
&\
+\sigma^{-2} \, M_{ij} \Big(
	Y^{I k} Y^{J l} \omega_{(a| k}{}^{i} \omega_{|b) l}{}^{j}
	- 2 A^{K}{}_{(a}  Y^{L i} Y^{I k}  \omega_{b) k}\,^{j} g_{K L}{}^{ J} 
%	\right.
\nonumber\\[3pt]
&
\qquad\qquad\qquad
	+A^{K}{}_{(a} A^{K'}{}_{b)}  Y^{L i} Y^{L' j} g_{K L}{}^I g_{K'L'}{}^J 	
\nonumber\\[3pt]
&
\qquad\qquad\qquad
	- 2 B_{(a| k} A^{K}{}_{|b)} Y^{K' k} Y^{L i} Y^{L' j} g_{KL }{}^{I}  g_{K'L'}{}^{J} 
\nonumber\\[3pt]
&
\qquad\qquad\qquad
	+B_{(a| k} B_{|b) l}    Y^{K k} Y^{K' l} Y^{L i} Y^{L' j}  g_{KL}{}^{I} g_{K' L'}{}^{J} 	
\nonumber\\[3pt]
&
\qquad\qquad\qquad
\left.
	-{\tfrac{1}{4}} \, Y^{Ii}Y^{Jk}\left(  \eta_a\, \eta_b\, \delta^j{}_k
+4 \eta_{(a}\, (\kappa_{b)})_{k}{}^j \right)
\right)
\ .
\label{eq:D7M2ab}
\end{align}
On the other hand, the bulk contribution was already studied in \cite{Dibitetto:2019odu} and has been reviewed in Appendix~\ref{app:bulkreduction}. As emphasized in the previous section, we now have to take into account that open strings backreact onto the bulk fields modifying their field strengths as follows,
\begin{align}
H_{abi} 
= & \ 
\overline{H}_{abi} \,-\, 2\left(\kappa_{[a}\right)_{i}{}^{j}B_{b]j} \,+\, \eta_{[a}B_{b]i}\ ,
\\[1mm]
F_{a} 
= & \  
\overline{F}_{a} +  \frac{\tilde\lambda_7^2}{2}\,\left(
	 \left(\kappa_{a}\right)_{i}{}^{j}\,Y^{Ii}\,Y^{Ik}\,\epsilon_{jk} 
	 -  g_{IJK} \cA^I{}_a Y^{Ji} Y^{Kj} \, \epsilon_{ij}
	 \right)\ ,
\label{eq:D7F1mod}
\\[1mm]
 F_{abi} =&\ 
\overline{F}_{abi}
-\, 2\left(\kappa_{[a}\right)_{i}{}^{j}C_{b]j} \,+\, \eta_{[a}C_{b]i}
+2 {F}_{[a} B_{b]i} 
-\overline{H}_{abi}  \chi 
%-\epsilon_{\rm D7} \, \tilde\lambda_7^2 \, \kappa_{[a|k}{}^j  \cA^{I}{}_{|b]}\, Y^{Ik}\,  \epsilon_{ji}
%\nonumber\\
%&\
%+\epsilon_{\rm D7} \, \tilde\lambda_7^2 \left( 
%	\cF^I{}_{ab}
%	+{\cal A}^{I}{}_{[a}\,\eta_{b]}
%	+\frac12 \overline{H}_{abj} Y^{Ij}  
%	\right) Y^{Ik} \epsilon_{ki}
\nonumber\\
&\
+\tilde\lambda_7^2 \left[
	\left(
	\cF^I{}_{ab}
	+{\cal A}^{I}{}_{[a}\,\eta_{b]}
	+\tfrac12 \overline{H}_{abj} Y^{Ij}  
	\right) \epsilon_{ki}
	+ \cA^{I}{}_{[a} \,(\kappa_{b]})_{k}{}^j  \,  \epsilon_{ji}
	\right]Y^{Ik}
	\label{eq:D7F3mod}
\ ,
\end{align}
where $\tilde \lambda_7\equiv (2\kappa^2_8 \, N_{\rm D7}\, T_{\text{D}7})^{1/2}\, \lambda$ and $g_{IJK}\equiv g_{IJ}{}^L\, \kappa_{LK}$. Details on the derivation of these modified field strengths are provided in Appendix~\ref{app:O7D7}.

All these partial results already allow us to compute the scalar potential. It turns out to be given by
\begin{align}
V_{\text{O7/D7}}
=&\
\frac14 \rho^{-1}\, \tau^{-2}\,  \omega_{mn}{}^r\, \omega_{pq}{}^s M^{nq} \left(
	M_{rs} 	\,M^{mp}  
	+2 \,\delta^m{}_s \, \delta^p{}_r
	\right)
+\frac12 \,\rho\,\tau^{-6}\,\sigma^{-2} \, F_a \, F_b\, M^{ab}
\nonumber\\
&\
+\frac14\,\sigma^{-2} \left(
	\rho^{-3}\,\tau^{-2} \,H_{abi}\, H_{cdj}
	+\rho^{-1}\,\tau^{-6} \,F_{abi}\, F_{cdj}
	\right)\,M^{ac} M^{bd} M^{ij}
\nonumber\\
&\
+ \tau^{-4}\, \sigma^2 \, (2\kappa_{8}^2) T_{\text{D}7}(N_{\rm D7}+8\, \epsilon_{\text{O}7})
\nonumber\\
&\
+\frac14\, \tilde \lambda_7^2\, \tau^{-4}\sigma^{-2}\, \kappa_{IJ}  \left(
	 2\, \sigma^{2} \, \mathbb{M}^{(2)IJ}{}_{ab}  \, M^{ab} 
	-\rho^{-2} \, \mathbb{M}^{(1)I}{}_{ab}  \, M^{bc} \, \mathbb{M}^{(1)J}{}_{cd}  \, M^{da} 
\right.
\nonumber\\
&
\left.
\qquad\qquad\qquad\qquad\qquad\qquad
+\rho^2 M_{ij} M_{i'j'} Y^{Ki} Y^{Li'} Y^{K'j} Y^{L'j'} g_{KL}{}^I  g_{K'L'}{}^{J} 
\right)
\ .
\label{eq:VD7O7-comp}
\end{align}
Let us stress three relevant aspects of the potential: (i) as expected from the tadpole condition, the term proportional to the (non-vanishing) effective tension is present, (ii) the last two lines arise from the DBI action of the D7 branes, and (iii) the WZ contributions are entirely encoded in the modified field strengths $F_{(1)}$ and $F_{(3)}$.\footnote{It is unclear to us the origin of the terms involving $\eta_a$ in the last lines of \eqref{eq:D7M1ab} and \eqref{eq:D7M2ab}. Nevertheless, as we are interested in configurations with non-Abelian fluxes, one of the quadratic constraints in \eqref{eq:QCD7O7} forces $\eta_a=0$.}

\subsection*{Matching with $\mathcal{N}=(1,1)$ gauged supergravity}
Since the O$7$ reduction Ansatz respects $\mathrm{SL}(2,\mathbb{R})_{a}\times\mathrm{SL}(2,\mathbb{R})_{i}\times G_{\mathrm{YM}}$ covariance, the fundamental index $M$ of $\mathrm{SO}(4,4+\nv)$ is split into
\be
\begin{array}{lclclclclclc}
M & \longrightarrow & a & \oplus & i & \oplus & \bar{a} & \oplus & \bar{i} & \oplus & I & ,
\end{array}\notag
\ee
where $a$ \& $\bar{a}$ ($i$ \& $\bar{i}$) are (anti)fundamental indices of $\mathrm{SL}(2,\mathbb{R})_{a(i)}$, while $I$ is an adjoint index of $G_{\mathrm{YM}}$.
The dictionary between flux parameters and embedding tensor components is summarized in Table \ref{Table:O7_Fluxes/ET}.
\begin{table}[http!]
\renewcommand{\arraystretch}{1}
\begin{center}
\scalebox{1}[1]{
\begin{tabular}{|c|c|c|}
\hline
IIB fluxes & $\Theta$ components &  Dictionary \\
\hline \hline
$\overline{H}_{abi}=\,\epsilon_{ab}\,h_{i}$ & $f_{ab\bar{i}}$ & $f_{ab\bar{i}}=\epsilon_{ab}\epsilon^{ij}h_{j}$ \\
\hline
$\omega_{ai}{}^{j}=\,\left(\kappa_a\right)_i{}^j+\frac{1}{2}\eta_a\delta^j_i$ & $f_{ai\bar{j}}\,\oplus\,\xi_{a}$ & $\begin{array}{lcl}f_{ai\bar{j}}&=&\left(\kappa_a\right)_i{}^j\\\xi_{a} &=& -\eta_a \end{array}$ \\
\hline
$\omega_{ab}{}^{c}=\,-2\eta_{[a}\delta^{c}_{b]}$ & $f_{ab\bar{c}}$ & $f_{ab\bar{c}}=\frac{1}{2}\epsilon_{ab}\epsilon^{cd}\eta_{d}$ \\
\hline
$\omega_{ij}{}^{a}=\,\epsilon_{ij}\theta^a$ & $\zeta_a$ & $\zeta_a=-\epsilon_{ab}\theta^{b}$ \\
\hline
$\overline{F}_{a}=\,f_{a}$ & $f_{aij}$ & $f_{aij}=\,\epsilon_{ij}f_{a}$ \\
\hline
$\overline{F}_{abi}=\,\epsilon_{ab}f_{i}$ & $f_{abi}$ & $f_{abi}=\, \epsilon_{ab}f_{i}$ \\
\hline
$\overline{\mathcal{F}}^{I}{}_{ab}=\,\epsilon_{ab}g^{I}$ & $f_{abI}$ & $f_{abI}=\,\tilde\lambda_7\epsilon_{ab}\delta_{IJ}g^{J}$ \\
\hline
$g_{IJ}{}^{K}$ & $f_{IJK}$ & $f_{IJK}=\, -\tilde\lambda_7^{-1}\,g_{IJ}{}^L \, \kappa_{LK}$ \\
\hline
\end{tabular}
}
\end{center}
\caption{\it The embedding tensor/fluxes dictionary for type IIB reductions with spacetime filling O7-planes and D7-branes. Note that $\kappa_a$ satisfies $\left(\kappa_a\right)_i{}^i=0$, and $\eta_a$ parametrizes the partial traces of $\omega$, while still respecting $\omega_{mn}{}^{n}=0$, which is required by unimodularity.} \label{Table:O7_Fluxes/ET}
\end{table}

With these non-vanishing components of the embedding tensor, the general form of the QC \eqref{Full_QC} reduces to 
\be\label{eq:QCD7O7}
\begin{array}{rcccccrc}
\epsilon^{ab}\,\left(\eta_{a}\left(\kappa_{b}\right)_{i}{}^{j}\,-\,\left(\kappa_{a}\right)_{i}{}^{k}\left(\kappa_{b}\right)_{k}{}^{j}\right) \, = \, 0 & , & & & & & \epsilon^{ab}\, f_a\eta_b \, = \, 0 & , \\[1mm]
\eta_{a}\,g_{IJK} \, = \, 0 & , & & & & & g^Ig_{IJK} \, = \, 0 & , \\[1mm]
g_{[IJ}{}^{I'}\,g_{K]LI'}  \, = \, 0 & , & & & & & \eta_{a}\theta^{b}\,-\,\frac{1}{2}\,\delta^{b}_{a}\eta_c\theta^c \, = \, 0 & , 
\end{array}
\ee
which exactly reproduce the consistency constraints coming from both the Bianchi and Jacobi identities of the corresponding flux background.  In particular, the first and last quadratic constraints arise from the Jacobi identities of the structure constants \eqref{eq:Jacobi-omega} associated to the group manifold.

For the evaluation of the general supergravity scalar potential \eqref{Full_VSugra} in this specific case, we again need to identify how $\Sigma$ and the coset representative $\mathcal{V}_{M}{}^{\underline{M}}$ are expressed as a function of the $17+4\nv$ scalars $\left(\rho,\tau,\sigma,M_{ab},M_{ij},B_{ai},C_{ai},\chi, \psi,\mathcal{A}^{I}{}_{a},Y^{Ii}\right)$.
We find
\be
\Sigma \, = \, \sigma \ ,
\ee
while
\be
\mathcal{V}_{M}{}^{\underline{M}} \, = \, \left(
\begin{array}{c|c|c}
\tau L_{m}{}^{\underline{m}} & \tau^{-1}\mathcal{C}_{mn}L^{n}{}_{\underline{m}}  & -A^{I}{}_{m}\delta^{\underline{I}}_{I} \\ \hline
\mathbb{O}_{4} & \tau^{-1}\,L^{m}{}_{\underline{m}} & \mathbb{O}_{4,N} \\ \hline
\mathbb{O}_{N,4} & \tau^{-1}\delta_{IJ}A^{J}{}_{n}L^{n}{}_{\underline{m}}  & \delta^{\underline{I}}_{I} 
\end{array}
\right) \ ,
\ee
where 
\be
L^{m}{}_{\underline{m}} \, \equiv \, \left(
\begin{array}{c|c}
\rho^{-1/2}\ell^{a}{}_{\underline{a}} &  \mathbb{O}_{2}\\ \hline
- \rho^{-1/2} B_{bj}\epsilon^{ji}\ell^{b}{}_{\underline{a}} & \rho^{1/2}\ell^{i}{}_{\underline{i}}
\end{array}
\right) \ ,
\ee
with $\ell^{a}{}_{\underline{a}}$ \& $\ell^{i}{}_{\underline{i}}$ satisfying $\ell^{a}{}_{\underline{a}}\ell^{b}{}_{\underline{a}}=M^{ab}$ and $\ell^{i}{}_{\underline{i}}\ell^{j}{}_{\underline{i}}=M^{ij}$, respectively.
The matrix $\mathcal{C}$ is defined as $\mathcal{C}_{mn} \equiv\gamma_{mn}-\frac{1}{2}A^{I}{}_{m}A^{I}{}_{n}$, in terms of
\be
\begin{array}{lclc}
\gamma_{mn} \, \equiv \, \left( \begin{array}{c  |  c} \begin{array}{cc}0 & \psi\\ -\psi & 0\end{array} & - C_{aj}\\ \hline C_{bi} & \begin{array}{cc}0 & -\chi\\ \chi & 0\end{array}\end{array}\right) \ & ,  \qquad & A^{I}{}_{m}\, \equiv \, \tilde\lambda_7\left(\begin{array}{c  | c}\mathcal{A}^{I}{}_{a} \ & \ Y^{Ij}\epsilon_{ji}\end{array} \right)  & .
\end{array}
\ee

Thus, plugging $\cV_M{}^{\underline{M}}$ together with the parametrization of the embedding tensor $\Theta\equiv\{\, f_{MNP}, \xi_M, \zeta_M\, \}$ in the gauged supergravity potential \eqref{Full_VSugra}, we obtain the same scalar potential as the one calculated from the compactification in \eqref{eq:VD7O7-comp}.

\section{O$9$/D$9$ \& Open Strings}
\label{section:O9}
Let us finally study type IIB reductions with spacetime filling O$9$-planes, \emph{i.e.} type I reductions. The orientifold planes in this case fill the entire ten-dimensional spacetime:
\be
\mathrm{O}9 \ : \quad
\underbrace{\times \ \times \ \times \ \times \ \times \ \times}_{\textrm{6D spacetime}} \ | \ \underbrace{\times \ \times \ \times \ \times}_{y^{m}}  \ , 
\notag
\ee
where $\sigma_{\mathrm{O}9}$ acts trivially on all the coordinates, due to the absence of transverse directions. In this case the $\mathbb{Z}_2$ action realizing the truncation is purely given by the worldsheet parity operator $\Omega$. The set of resulting scalar modes retained by this operation is shown in Table \ref{Table:O9_Projection}.
\begin{table}[http!]
\renewcommand{\arraystretch}{1}
\begin{center}
\scalebox{1}[1]{
\begin{tabular}{|c|c|c|c|}
\hline
IIB fields & $\sigma_{\mathrm{O}9}$ & $\Omega$ & \# dof's   \\
\hline \hline
$e^{m}{}_{n}$ & $+$ & $+$ & $16-6\,=\,10$ \\
\hline 
$B_{mn}$ & $+$ & $-$ & --- \\
\hline
$\Phi$ & $+$ & $+$ & $1$ \\
\hline \hline
$C_{(0)}$ & $+$ & $-$ & --- \\
\hline 
$C_{mn}$ & $+$ & $+$ & $6$ \\
\hline 
$C_{mnpq}$ & $+$ & $-$ & --- \\
\hline \hline
$\mathcal{A}^{I}{}_{m}$ & $+$ & $+$ & $4\nv$ \\
\hline
\end{tabular}
}
\end{center}
\caption{\it The $\mathbb{Z}_{2}$ parity of all internal components of the different IIB fields in the presence of spacetime filling O$9$-planes. The allowed ones yield excitable 6D scalar fields. Note that the total amount of resulting scalars correctly gives $17+4\nv$, \emph{i.e.} the dimension of the supergravity coset \protect\eqref{G_global_coset}.} \label{Table:O9_Projection}
\end{table}

The spacetime filling O$9$-planes preserve the internal diffeomorphism covariance, thus making the reduction scheme to be formulated in a $\mathrm{SL}(4,\mathbb{R})\times G_{\mathrm{YM}}$ covariant way. The ten-dimensional supergravity Ansatz that contains the $17$ closed string scalars coincides with the one studied by Kaloper and Myers in \cite{Kaloper:1999yr} except for the fact that they work in the heterotic frame and focus on the Abelian case.\footnote{The compactification of the heterotic string effective action on a torus including the non-Abelian sector has been done in \cite{Ortin:2020xdm}.} In terms of the type I fields, our Ansatz reads 
\begin{align}
\begin{split}
ds_{(10)}^2=&\ \tau^{-2}\,g_{\mu\nu}dx^{\mu}dx^{\nu} \, 
+\, \rho \, M_{m n} \, v^{m}v^{n}+\dots\ , 
\\[1mm]
e^{\Phi}=&\ \rho\,\tau^{-2}\ , 
\\[1mm]
C_{(2)} = &\ \frac12 \, C_{mn}  \, v^m \wedge v^n  - \frac{\tilde\lambda_9^2}{2} {\cal A}^{I}\wedge \sigma^I+\gamma\, + \, \dots \ , 
\end{split}
\label{eq:D9-ansatz}
\end{align}
where $\tilde\lambda_9\equiv (2\kappa_{10}^2\,N_{\rm D9}\, T_{\textrm{D}9}\lambda^2)^{1/2}$.  Once again, the scalars $\rho$ and $\tau$ represent the volume and dilaton would-be moduli, $M_{mn}$ is an element of $\mathrm{SL}(4,\mathbb{R})/\mathrm{SO}(4)$, and $C_{mn}$ parametrizes the axionic scalars coming from the R-R 2-form $C_{(2)}$. As expected, these modes add up to $17$. As before, we have introduced a parallelization of the internal manifold with torsion given in terms of the Maurer-Cartan one-forms $v^{m}$, which fulfill the equation
\be
dv^{m}\,+\,\frac{1}{2}\,\omega_{np}{}^{m}v^{n}\wedge v^{p} \, = \, 0 \ .
\ee

In addition to these fields, when adding a stack of $N_{\Dp9}$ D$9$-branes parallel to the O$9$, we will require $\nv$ extra vector multiplets, labelled by the index $I$ to accommodate the full group $G_{\rm YM}$. In terms of 6D degrees of freedom, besides the new vector fields, our resulting gauged supergravity description will again have $4\nv$ new scalar modes arising from the internal components of the worldvolume gauge fields
\be
\mathcal{A}^{I} \, =\, \mathcal{A}^{I}{}_{m}v^{m}  + \sigma^I+ \ldots\ .
\label{eq:D9-ansatz2}
\ee
The 1- and 2-form $\sigma^I(y)$ and $\gamma(y)$ in the Ansatze \eqref{eq:D9-ansatz} and \eqref{eq:D9-ansatz2} give rise to the vector flux $\overline \cF^{I}{}_{mn}$ and the 3-form flux $\overline F_{mnp}$ listed in Table \ref{Table:O9_Fluxes/ET} via their exterior derivatives \cite{Kaloper:1999yr},
\begin{eqnarray}
d\sigma^{I}-\frac{1}{2}\overline \cF^{I}{}_{mn}v^m\wedge v^n&=&0\,,\\[1mm]
d\gamma- \frac{\tilde\lambda_9^2}{2}\sigma^I\wedge d\sigma^I&=&\frac{1}{3!}\overline F_{mnp}\,v^m\wedge v^n \wedge v^p\,.
\end{eqnarray}
The integrability conditions associated to the above equations are,
\begin{eqnarray}
\overline \cF^{I}{}_{q[m}\, \omega_{np]}{}^{q}&=&0\,,\\[1mm]
 \frac{\tilde\lambda_9^2}{2}\, \overline \cF^I{}_{[mn}\, \overline \cF^I{}_{pq]}-\overline F_{r[mn}\, \omega_{pq]}{}^{r}&=&0\,.
\end{eqnarray}

In order to compute the contribution to the scalar potential coming from the DBI action \eqref{eq:DBI-Dbrane}, we just need to know explicitly the matrix ${\mathbb M}_{MN}$ \eqref{eq:matrixM} since the lack of transverse coordinates forbids the existence of $\mathbb{Q}$. The non-trivial components of ${\mathbb M}_{MN}$ are 
\begin{align}
\mathbb{M}_{\mu\nu} = & \ \tau^{-2}\,g_{\mu\nu} \,+\, \ldots \ ,\\[1mm]
\mathbb{M}_{mn} = & \ \rho\,M_{mn} + \lambda\, \cF_{mn}  \,+\, \ldots\ .
\end{align}
Instead, the bulk contribution can be extracted from using the results provided in Appendix~\ref{app:bulkreduction}, but we need to compute first $F_{mnp}$. To this aim, one has to bear in mind that the coupling of the sources to $C_{(6)}$ modifies the Bianchi identity of $F_{(3)}$ as in \eqref{eq:BianchiF3}, which we repeat here for convenience
\begin{equation}
dF_{(3)}=-\frac{{\tilde\lambda}_9^2}{2}{\cal F}^{I}\wedge {\cal F}_{I}\,.
\end{equation}
This implies that, locally, the field strength $F_{(3)}$ is given by 
\begin{equation}
F_{(3)}=\,dC_{(2)}-\frac{{\tilde\lambda}^2_9}{2}\left(\cF^I\wedge \cA^I+\frac{1}{3!}\,g_{IJK}\, \cA^I\wedge \cA^J\wedge \cA^K\right)\, .
\end{equation}
Restricting to the internal components components and making use of the reduction Ansatz, \eqref{eq:D9-ansatz} and \eqref{eq:D9-ansatz2}, we obtain that
\begin{multline}
F_{mnp}
=
	\overline F_{mnp}
-3 \, C_{q[m}\,\omega_{np]}{}^q
-3 \, \tilde \lambda_9^2 \, {\cal A}^I{}_{[m|} \left(
	\overline \cF^I{}_{|np]}	
	-\frac13\,  g_{IJK}  {\cal A}^J{}_{|n} {\cal A}^K{}_{p]}
	-\frac12{\cal A}^I{}_q \, \omega_{|np]}{}^q
\right)
\ .
\end{multline}

Finally, the complete expression for the scalar potential is,
\begin{align}
\begin{split}
V_{\textrm{O}9/\textrm{D}9} =&\ 
\frac14 \rho^{-1} \tau^{-2}  \left(
	M_{m n} M^{p q} M^{rs} \omega_{p r}{}^{m} \omega_{q s}{}^{n}
    +2M^{m n} \omega_{m q}{}^{p} \omega_{n p}{}^{q} 
    \right)
\\[3pt]
& 
+\frac{\tilde\lambda_9^2}{4} \rho^{-1} \tau^{-4} M^{m m'} M^{n n'} \cF^{I}{}_{m n} \cF^{I}{}_{m' n'} 
+ \frac{1}{12} \rho^{-1} \tau^{-6} M^{m m'} M^{n n'} M^{p p'} F_{mnp} F_{m'n'p'}
\\[3pt]
&
+2\, \kappa_6^2\, T_{\text{D}9}(N_{\Dp9}+32\, \epsilon_{\text{O}9})\rho\tau^{-4}
\ ,
\end{split}
\label{eq:V-D9}
\end{align}
where
\begin{align}
\begin{split}
\cF^{I}{}_{mn}
\equiv &\
	\overline \cF^{I}{}_{m n} 
	-g_{JK}{}^{I} \cA^{J}{}_{m} \cA^{K}{}_{n} 
	-\cA^{I}{}_{p} \, \omega_{m n}{}^{p}
\ .
\end{split}
\end{align}
As a tadpole cannot be generated with the fluxes at our disposal, the condition $N_{\Dp9}+32\, \epsilon_{\text{O}9}=0$ must be imposed. This implies $N_{\Dp9}=32$ and $\epsilon_{\rm O9}=-1$, so that the total charge and tension vanish. Therefore, the last term in the scalar potential, which is proportional to the effective tension, vanishes as well.

\subsection*{Matching with $\mathcal{N}=(1,1)$ gauged supergravity}
The O$9$ reduction Ansatz preserves $\mathrm{SL}(4,\mathbb{R})\times G_{\mathrm{YM}}$ covariance, in such a way that the fundamental index $M$ of $\mathrm{SO}(4,4+\nv)$ is split into
\be
\begin{array}{lclclclclclc}
M & \longrightarrow & m & \oplus & \bar{m} & \oplus & I & ,
\end{array}\notag
\ee
where $m$ \& $\bar{m}$ are (anti)fundamental indices of $\mathrm{SL}(4,\mathbb{R})$, and $I$ is an adjoint index of $G_{\mathrm{YM}}$.
The dictionary between flux parameters and embedding tensor components is summarized in Table \ref{Table:O9_Fluxes/ET}.
\begin{table}[http!]
\renewcommand{\arraystretch}{1}
\begin{center}
\scalebox{1}[1]{
\begin{tabular}{|c|c|c|}
\hline
IIB fluxes & $\Theta$ components &  Dictionary \\
\hline \hline
$\overline F_{mnp}$ & $f_{mnp}$ & $f_{mnp}=\, -\overline F_{mnp}$ \\
\hline
$\omega_{mn}{}^{p}$ & $f_{mn\bar p}$ & $f_{mn\bar p}=\omega_{mn}{}^p$ \\
\hline
$\overline \cF^{I}{}_{m n}$ & $f_{m n I}$ & $f_{m n I}=-\tilde\lambda_9\, \kappa_{IJ} \, \overline \cF^{J}{}_{m n}$ \\
\hline
$g_{IJ}{}^{K}$ & $f_{IJK}$ & $f_{IJK}= \tilde \lambda_9^{-1} \,g_{IJ}{}^{L}\kappa_{LK}$ \\
\hline
\end{tabular}
}
\end{center}
\caption{\it The embedding tensor/fluxes dictionary for type IIB reductions with spacetime filling O9-planes and D9-branes. Note that $\omega_{m n}{}^p$ is restricted to satisfy $\omega_{mn}{}^n=0$, which is required for unimodular gaugings.
} \label{Table:O9_Fluxes/ET}
\end{table}

With these non-vanishing components of the embedding tensor, the general form of the QC \eqref{Full_QC} reduces to 
\be
\begin{array}{rcccccrc}
-\overline F_{r[mn}\, \omega_{pq]}{}^r + \dfrac{\tilde\lambda_9^2}{2} \, \overline \cF^I{}_{[mn} \, \overline\cF^{I}{}_{pq]}\, = \, 0 & , 
& & & & & 
\omega_{[mn}{}^r\, \omega_{p]r}{}^q  \, = \, 0 & , 
\\[1mm]
\omega_{[mn}{}^q\, \overline \cF^I{}_{p]q} \, = \, 0 & , 
& & & & & 
g_{IJK} \, \overline \cF^K{}_{mn} \, = \, 0 & , \\[1mm]
g_{[IJ}{}^{I'}\,g_{K]LI'}  \, = \, 0 & , 
\end{array}
\label{eq:D9-QC}
\ee
which exactly reproduce the consistency constraints coming from the BI of the corresponding flux background (see Appendix \ref{app:O9D9} for further details).

For the evaluation of the general supergravity scalar potential \eqref{Full_VSugra} in this specific case, we again need to identify how $\Sigma$ and the coset representative $\mathcal{V}_{M}{}^{\underline{M}}$ are expressed as a function of the $17+4\nv$ scalars. In this case, because no fluxes are embedded into $\zeta_M$, the scalar potential is entirely written in terms of ${\cal H}_{MN}$.
We find then
\be
\Sigma \, = \, \rho^{1/2} \ , \qquad\qquad \Lambda \, = \, \tau^2 \ ,
\ee
while
%
%\be
%\mathcal{H}_{MN} \, = \, \left(
%\begin{array}{ccc}
%\Lambda\, M_{m n}+\Lambda^{-1}\, {\cal C}_{p m} \,{\cal C}_{q n} \,M^{m n} +  \kappa_{IJ}\,  \mathscr{A}^I{}_{m} \,\mathscr{A}^J{}_n 
%& 
%-\Lambda^{-1} \,M^{np} \,{\cal C}_{p m}  
%& 
%\mathscr{A}^{J}{}_m + \Lambda^{-1} \,{\cal C}_{p m} \,M^{pq} \,\mathscr{A}^{J}{}_q 
%\\[.5em]
%-\Lambda^{-1}\, M^{mp}\,{\cal C}_{pn}
%& 
%\Lambda^{-1} \, M^{mn}
%& 
%-\Lambda^{-1} M^{mp} \,\mathscr{A}^{J}{}_{p}
%\\[.5em]
%\mathscr{A}^{I}{}_n+\Lambda^{-1}\, {\cal C}_{pn} \, M^{pq} \, \mathscr{A}^I{}_q
%& 
%-\Lambda^{-1} M^{np} \,\mathscr{A}^I{}_p
%& 
%\kappa_{IJ}+\Lambda^{-1}\, \mathscr{A}^I{}_p\, \mathscr{A}^J{}_q\, M^{pq}
%\end{array}
%\right) \ ,
%\ee
\be
\mathcal{H}_{MN} \, = \, \left(
\begin{array}{ccc}
\tilde M_{m n}+ {\cal C}_{p m} \,{\cal C}_{q n} \,\tilde M^{pq} +  \kappa_{IJ}\,  A^I{}_{m} \,{A}^J{}_n 
& 
-\tilde M^{np} \,{\cal C}_{p m}  
& 
{A}^{J}{}_m +  \,{\cal C}_{p m} \,\tilde M^{pq} \, {A}^{J}{}_q 
\\[.6em]
-\tilde M^{mp}\,{\cal C}_{pn}
& 
\tilde  M^{mn}
& 
-\tilde  M^{mp} \,{A}^{J}{}_{p}
\\[.6em]
{A}^{I}{}_n+ {\cal C}_{pn} \, \tilde M^{pq} \, {A}^I{}_q
& 
- \tilde M^{np} \,{A}^I{}_p
& 
\kappa_{IJ}+ {A}^I{}_p\, {A}^J{}_q\, \tilde M^{pq}
\end{array}
\right) \ ,
\ee
where 
\begin{align}
\tilde M_{mn} \equiv \Lambda \, M_{mn} 
\ ,
\qquad
{\cal C}_{mn} 
\equiv \, C_{mn}
+\frac12 \kappa_{IJ} \,{A}^I{}_m \, {A}^J{}_n
\ ,
\qquad
{A}^I{}_m \equiv \tilde  \lambda_9\, \cA^I{}_m
\ .
\end{align}

\section*{Concluding Remarks}

In this paper we have studied type IIB flux compactifications down to six dimensions with spacetime filling O-planes, D-branes and open strings. Such compactifications turn out to yield 6D $\mathcal{N}=(1,1)$ gauged supergravities. The exact relation between the 10D \& the 6D descriptions was studied in \cite{Dibitetto:2019odu} within the closed string sector. Now we have been able to generalize this correspondence to the case where open strings are excited. We also included open string effects such as non-Abelian D-brane gauge groups and non-trivial YM flux. Our analysis is very much in the spirit of the one in \cite{Aldazabal:2008zza} carried out in the context of compactifications down to four dimensions, the main difference being that our brane gauge groups may be non-Abelian.

The dictionary derived here allowed us to understand crucial physical mechanisms like the bulk field strength modification induced by open string effects. At least at a heuristic level, this can be related to the GS mechanism for the heterotic string via a duality chain. Besides this intuition, we were able to explain such a form of the bulk field strengths directly in terms of couplings contained in the WZ brane actions.

The present work sets the ground for interesting developments within the context of compactifications of type IIB string theory with sixteen supercharges. This will first of all, allow us to search for string vacua supported by interactions between the open and the closed string sectors.
This moves towards the direction of  \cite{Escobar:2018tiu,EscobarAtienzar:2019zxp,Herraez:2018vae}, where a similar analysis was performed in type IIA strings and conditions for the existence of 4D vacua with \emph{mobile} D-branes are discussed. Our machinery could be extremely valuable in improving efficiency for vacua searches.

Moreover, the possible existence of certain types of vacua within this setup could shed a light on issues of utmost importance, like the validity of Swampland conjectures or the string universality principle. To this end, it would be very interesting to explore the set of non-supersymmetric vacua of this sort (AdS or dS, even), or to study those consistency conditions for flux backgrounds coming from anomaly cancellation. Finally, if one could find supersymmetric AdS extrema in this class, it would be interesting to study their holographic interpretation.
We certainly intend to address all of these issues in the next future.

\section*{Acknowledgements}

The work of JRB is supported by Fundación S\'eneca, Agencia de Ciencia y Tecnolog\'ia de la Regi\'on de Murcia under grant 21472/FPI/20.
The work of GD is supported by the STARS grant named THEsPIAN.
The work of JRB and JJFM is supported by the Spanish Ministerio de Innovaci\'on y Ciencia, CARM Fundaci\'on S\'eneca and Universidad de Murcia under grants PID2021-125700NA-C22, 21257/PI/19 and E024-18, respectively. AR is supported by the MIUR-PRIN contract 2020KR4KN2  ``String Theory as a bridge between Gauge Theories and Quantum Gravity". JRB and JJFM acknowledge  Universit\`a di Padova for hospitality while this work was carried out.

\appendix
\section{Type IIB Supergravity in the String Frame}
\label{app:iib-string}

Type IIB supergravity in its democratic formulation \cite{Bergshoeff:2001pv} is described in terms of the common NS-NS sector $\{G, B_{(2)},\Phi\}$ coupled to even form fields $\{C_{(2p)}\}_{p=0,1,2,3,4}$. The (bosonic) dynamics of the theory can be derived from the following \emph{pseudoaction}
\begin{equation}
\label{IIB_action}
S_{\mathrm{IIB}} \, = \, \frac{1}{2\kappa_{10}^2}\int{d^{10}x\sqrt{-G}\left(e^{-2\Phi}\left(\mathcal{R}+4(\partial\Phi)^2-\frac{1}{12}|H_{(3)}|^2\right)-\frac{1}{4}\sum\limits_{p=0}^{4}\frac{|F_{(2p+1)}|^2}{(2p+1)!}\right)} \ ,
\end{equation}
where the field strengths read
\begin{align}
\begin{split}
H_{(3)} =&\ dB_{(2)}\ , \\
H_{(7)}=&\ dB_{(6)}+\frac{1}{2}(-F_{(7)}\wedge C_{(0)}+F_{(5)}\wedge C_{(2)}-F_{(3)}\wedge C_{(4)}+F_{(1)}\wedge C_{(6)})\ , \\
F_{(1)}=&\ dC_{(0)} \ , \\ 
F_{(3)} =&\  dC_{(2)}-H_{(3)}\wedge C_{(0)} \ , \\
F_{(5)}=&\ dC_{(4)}-H_{(3)}\wedge C_{(2)}\ ,\\
F_{(7)}=&\ dC_{(6)}-H_{(3)}\wedge C_{(4)}\ ,\\
F_{(9)}=&\ dC_{(8)}-H_{(3)}\wedge C_{(6)}\ ,
\end{split}
\label{fieldstrenghts}
\end{align}
which are designed to automatically satisfy the following (modified) Bianchi identities (BI)
\begin{align}
dH_{(3)}=&\ 0\ , 
&
dH_{(7)}+\frac{1}{2} \sum\limits_p \star F_{(p)} \wedge F_{(p-2)} =&\ 0\ , \label{BIH3} 
\\[-2mm]
dF_{(1)}=&\ 0\ ,
&
dF_{(3)}-H_{(3)}\wedge F_{(1)}=&\ 0\ ,
\\
dF_{(5)}-H_{(3)}\wedge F_{(3)}=&\ 0\ ,
&
dF_{(7)}-H_{(3)}\wedge F_{(5)}=&\ 0\ ,
\\
dF_{(9)}-H_{(3)}\wedge F_{(7)}=&\ 0\ .\label{BIG9}
\end{align}
It is worth mentioning that $S_{\mathrm{IIB}}$ in \eqref{IIB_action} is called a pseudoaction because it must be supplemented by the following \emph{duality relations}
\begin{align}
F_{(9)}\overset{!}{=}\star F_{(1)}
\ ,
\quad
F_{(7)}\overset{!}{=}-\star F_{(3)}
\ ,
\quad
F_{(5)}\overset{!}{=}\star F_{(5)}
\ ,
\quad
\ H_{(7)}\overset{!}{=}e^{-2\Phi}\star H_{(3)}
\ ,
\label{eq:IIBdualities}
\end{align}
that yield the correct number of propagating degrees of freedom and hence allow for an on-shell realization of supersymmetry.
By varying  \eqref{IIB_action}, one obtains the following set of equations of motion
\begin{equation}\label{EOMPhi}
\Box\Phi \,-\, (\partial\Phi)^2\,+\,\frac{1}{4}\,\mathcal{R}\,-\,\frac{1}{8\times 3!}\,|H_{(3)}|^2\, = \, 0 \ ,
\end{equation}
for the 10D dilaton $\Phi$,
\begin{align}
 d(e^{-2\Phi}\star H_{(3)})+\frac{1}{2}\sum\limits_p \star F_{(p)}\wedge F_{(p-2)}=&\ 0\ , & d(e^{2\Phi}\star H_{(7)})= & \ 0\ , \\
 d(\star F_{(1)})+H_{(3)}\wedge(\star F_{(3)})=& \ 0\ , & d(\star F_{(3)})+H_{(3)}\wedge F_{(5)}=&\ 0\ ,\\[3mm]
 d(\star F_{(5)})- H_{(3)}\wedge F_{(3)} =&\ 0 \ ,
\end{align}
for the form fields, and finally the (trace reversed) Einstein equations
\begin{align}\label{EinsteinEQ}
0=& \ e^{-2\Phi}\left(
	\mathcal{R}_{\cM\cN}
	+2\nabla_\cM\nabla_\cN\Phi
	-\frac{1}{4} H_{\cal MPQ} H_\cN{}^{\cal PQ}
	\right)
-\frac{1}{2} (F_{(1)}^2)_{\cal MN} \notag\\
&-\frac{1}{2\times 2!} (F_{(3)}^2)_{\cal MN}
-\frac{1}{4\times 4!} (F_{(5)}^2)_{\cal MN}
+\frac{1}{4}G_{\cal MN}\left(
	|F_{(1)}|^2
	+\frac{1}{3!}|F_{(3)}|^2
	\right)
\ .
\end{align}

\section{Bulk Reduction}\label{app:bulkreduction}

In order to carry out the dimensional reduction of type IIB down to six dimensions, we parametrize the ten-dimensional metric $G_{{\cal M}{\cal N}}$ in terms of the six-dimensional one and the moduli describing the four-dimensional internal metric. In particular, by picking
\begin{align}
ds_{(10)}^2
=
G_{\cal MN}\, dx^\cM\otimes dx^\cN
=
\tau^{-2}\,  g_{\mu\nu}^{(6)} \, dx^\mu\otimes dx^\nu+\rho \, ds_{(4)}^2
\ ,
\end{align}
the universal moduli $\rho$ and $\tau$ are singled out, whereas the rest of the moduli are encoded inside $g^{(4)}$ and describe volume preserving deformations of the internal geometry. In addition to that, we introduce local indices $m$, $n$ as
\begin{align}
ds_{(4)}^2 = \cM_{mn} \, v^m \otimes v^{n}
\ ,
\end{align}
where the matrix $\cM_{mn}$ parametrizes the coset $\text{SL}(4,\mathbb{R})/\text{SO}(4)$ and $\det\cM=1$.

To obtain the 6D gravity action in the Einstein frame upon compactification, we require the constraint \cite{Hertzberg:2007wc} 
\begin{align}
\rho^2 \stackrel{!}{=}e^{2\Phi}\, \tau^4
\ ,
\end{align}
which implies that $\rho$ and $\tau$ fix the internal volume and the string coupling.

Let us consider now each of the terms in the type IIB effective action. The determinant of the metric reduces  to
\begin{align}
\sqrt{-G}
\quad \to\quad
\tau^{-6}\, \rho^2\, \sqrt{g_{(4)}}\, \sqrt{g_{(6)}}
\ . 
\end{align}

Then, the reduction of the Einstein term in \eqref{IIB_action} amounts to
\footnote{Please note that $\cR^{(10)}\, \to\, \tau^2\, \cR^{(6)}+\rho^{-1}\, \cR^{(4)}$. }
\begin{align}
\int \dd^{10} x\, \sqrt{-G} \, e^{-2\Phi}\cR^{(10)}
\quad \to&\  \quad
\int \dd^{10} x\, \sqrt{-g_{(6)}}\left(
	\tau^{-4}\, \rho^2\, e^{-2\Phi}\, \cR^{(6)}
	+\tau^{-6} \, \rho \, e^{-2\Phi}\, \cR^{(4)}
	\right)
\nonumber\\
=&\  \quad
\int \dd^{10} x\, \sqrt{-g_{(6)}}\left(
	  \cR^{(6)}
	-V_\omega
	\right)
\ ,
\end{align}
where $V_\omega\equiv -\rho^{-1}\, \tau^{-2}\, \cR^{(4)}$. In case of twisted toroidal compactifications, where $v^m$ are the Maurer-Cartan 1-forms, $\cR^{(4)}$ has the following expression \cite{Scherk:1979zr}:
\begin{align}
\cR^{(4)}
=
-\frac{1}{4}\, \cM_{mq} \cM^{nr} \cM^{ps} \, \omega_{np}{}^q \, \omega_{rs}{}^m
-\frac12 \, \cM^{np}\, \omega_{mn}{}^q\, \omega_{qp}{}^m
\ ,
\end{align}
where $\cM^{mn}$ is the inverse of $\cM_{mn}$ and $\omega_{mn}{}^p$ are the structure constants entering the Maurer-Cartan equation
\begin{align}
d v^m
+\frac12 \omega_{np}{}^m \, v^n \wedge v^p
= 0
\ . 
\end{align}
This, in turn, implies the Jacobi identities as an integrability condition,
\begin{align}
\omega_{[mn}{}^r\, \omega_{p]r}{}^q = 0
\ .
\end{align}
In addition to this, we will ask the structure constants to fulfill the unimodularity condition $\omega_{mn}{}^n=0$ for consistency, as we are performing the compactification at the level of the action.

Let us consider now the scalar potential arising from the $H$ flux. Reducing the corresponding term of the action \eqref{IIB_action} yields
\begin{align}
\int \dd^{10}x\, \sqrt{-G}\left(
	-\frac{1}{12}\, e^{-2\Phi}\, |H_{(3)}|^2
	\right)
\quad
\to
\quad
\int \dd^{6}x\, \sqrt{-g_{(6)}}\left(
	-\frac{1}{12}\, H_{mnp}\, H^{mnp} \, \rho^{-3}\, \tau^{-2}
	\right)
\ ,
\end{align}
so that the contribution consists of $V_H\equiv \frac{1}{12}\, H_{mnp}\, H^{mnp} \, \rho^{-3}\, \tau^{-2}$ and the contraction with the indices is done with the internal metric $g^{(4)}$.

Regarding the R-R $p$-forms\footnote{Please note that the numerical factor in \eqref{IIB_action} is $-\frac{1}{4p!}$ due to the simultaneous presence of the dual magnetic fields $F_{(10-p)}$. Using the duality relations \eqref{eq:IIBdualities}, the factor $\frac{1}{2p!}$ is trivially restored.}, their contribution to the scalar potential is
\begin{align}
\int \dd^{10} x \, \sqrt{-G}\left(
	-\frac{1}{2p!}\,  |F_{(p)}|^2
	\right)
\quad
\to
\quad
\int \dd^{6}x\, \sqrt{-g_{(6)}}\left(
	-\frac{1}{2p!}\, F_{m_1\cdots m_p}\, F^{m_1\cdots m_p} \, \rho^{2-p}\, \tau^{-6}
	\right)
\ ,
\end{align}
so that $V_{F_p}=\frac{1}{2p!}\, F_{m_1\cdots m_p}\, F^{m_1\cdots m_p} \, \rho^{2-p}\, \tau^{-6}$.

Thus, as the 10D Chern-Simons term does not give any contribution to the potential, the reduced $D=6$ theory is given by the following action
\begin{align}
S_{6\text{D}}
=
\int \dd^6x \sqrt{-g^{(6)}}\, \left(
	\cR^{(6)}
	+2\cL_{\text{kin}}
	-V
	\right)
\ , 
\end{align}
where the full scalar potential arising from the bulk and the effective tension consists of
\begin{align}
V
=
V_\omega+V_H+\sum_p V_{F_p}
\ .
\end{align}
The kinetic term for the moduli, which span a $\mathbb{R}^+_\rho\times \mathbb{R}^+_\tau
\times \text{SL}(4,\mathbb{R})/\text{SO}(4)$ geometry, is given by
\begin{align}
\cL_{\text{kin}}
=
-\frac{(\partial\rho)^2}{2\rho^2}
-\frac{(\partial\tau)^2}{\tau^2}
+\frac{1}{8}\Tr{\partial\cM\partial\cM^{-1}}
\ .
\end{align}

\section{Non-Abelian Brane Actions and Reductions Thereof}
\label{app:iib-DBI}
D$p$-brane actions on curved backgrounds with fluxes were studied in \cite{Myers:1999ps,Martucci:2005rb}. In this appendix we collect some relevant details for evaluating the bosonic effective actions of an O$p$-plane and a stack of $N_{\Dp p}$ D$p$-branes contributing to the scalar potential upon compactification. We will follow the notation in \cite{Choi:2018fqw}. As opposed to the rest of this paper\footnote{Note that this exact notation directly applies to the case of D$5$-branes, where the worldvolume exactly coincides with the 6D spacetime and all the internal directions are then transverse. However, in the other cases covered by our analysis, one has to suitably adapt the notation of this section before using the relations appearing here.}, we denote by $x^{M}$ the worldvolume coordinates, whereas the transverse coordinates to the branes are called $y^{i}$. Therefore, the 10D spacetime coordinates $x^{\cM}$ split into
$x^{\cM}=(x^M\, , \, y^i)$.

We will work in the \emph{static gauge}, where the position of each brane reads $y^{Ii}\,=\, \lambda\, Y^{Ii}$, with $\lambda\,\equiv\,2\pi\ell_s^2$. We will consider the generators of $G_{\textrm{YM}}$ to live in the fundamental representation of the Lie algebra. Denoted by $\{t_I\}_{I=1,\dots,\nv}$,
they satisfy
\be
\begin{array}{lccccl}
\left[t_{I},t_{J}\right] \,=\, i \, g_{IJ}{}^{K}\,t_{K} \ , & & & & & \textrm{Tr}(t_{I}t_{J}) \,=\, N_{{\rm D}p} \, \kappa_{IJ}\ ,
\end{array}
\ee
where $\kappa_{IJ}=g_{IK}{}^{L}g_{JL}{}^{K}$ is the Cartan-Killing metric of $G_{\textrm{YM}}$.

The bosonic worldvolume action describing a stack of $N_{\Dp p}$ coincident D$p$-branes in type II string theory contains two pieces: the Dirac-Born-Infeld (DBI) and the Wess-Zumino (WZ) actions,
\be
S_{\textrm{D}p} \ = \ S_{\textrm{D}p}^{\textrm{DBI}} \, + \, S_{\textrm{D}p}^{\textrm{WZ}} \ ,
\ee
where
\be
S_{\textrm{D}p}^{\textrm{DBI}} \, = \, -T_{\textrm{D}p}\int{d^{p+1}x\, \mathrm{Tr}\left(e^{-\hat{\Phi}}\sqrt{-\mathrm{det}(\mathbb{M}_{MN})\,\mathrm{det}(\mathbb{Q}^{i}{}_{j})}\right)} \ ,
\label{eq:DBI-Dbrane}
\ee
being $T_{\textrm{D}p}=2\pi \ell_s^{-(p+1)}$ the brane tension. The matrices $\mathbb{M}$ and $\mathbb{Q}$ are defined as
\begin{align}
\mathbb{M}_{MN}
=&\
\ \mathrm{P}\left[
	\hat E_{MN}
	+\hat E_{Mi}(\mathbb{Q}^{-1}-\delta)^{ij}\hat E_{jN}
	\right]
	+\lambda \mathcal{F}_{MN}
\ ,
\label{eq:matrixM}
\\
\mathbb{Q}^i{}_j
=&\
\delta^i{}_j +i \lambda [Y^i,Y^k] \hat E_{kj}
\ .
\label{eq:matrixQ}
\end{align}

On the other hand, the WZ action reads
\be
S_{\textrm{D}p}^{\textrm{WZ}} \, = \, \mu_{\textrm{D}p}\int\limits_{\textrm{WV}(\textrm{D}p)}{\, \mathrm{Tr}\left(\mathrm{P}\left[e^{i\lambda\iota_Y\iota_Y}\hat{C}\wedge e^{\hat{B}_{(2)}}\wedge e^{\lambda\mathcal{F}}\right]\right)} \ ,
\label{eq:WZ-action}
\ee
where $\mu_{{\Dp} p}=\epsilon_{\Dp p}\, T_{\Dp p}$, being $\epsilon_{\Dp p}=\pm1$ the charge sign corresponding to the brane and anti-brane cases respectively.

The field strength $\cF$ living on the brane is given by
\begin{align}
\cF = \dd \cA+ i\cA \wedge \cA 
\ ,
\end{align}
whereas
\begin{align}
\hat {E}_{\cM\cN} = \hat G_{\cM\cN} +\hat B_{\cM\cN}
\ .
\end{align}
The hat ``\^{}'' on the fields indicates that they are evaluated at the position of the D$p$-branes placed at $y^i=\lambda Y^i$, which is defined via a Taylor expansion as, for example, 
\begin{align}
\hat \phi(x^M,\lambda Y^i)
\equiv
\left.\sum_{n=0}^\infty \frac{\lambda^n}{n!} \, Y^{i_1}\, \cdots \, Y^{i_n}\, \partial_{i_1}\,\cdots\,\partial_{i_n}\,\phi(x^M,y^i)\right|_{y^i=0}
\label{eq:Taylor}
\ .
\end{align} 
With $\textrm{P}[\cdots]$ we denote the pullback of the bulk fields over the D$p$-brane worldvolume, in such a way the ordinary derivative $\partial_M Y^i$ is substituted by the covariant derivative $D_M Y^i$:
\begin{align}
D_M Y^i\equiv \partial_M Y^i+i[A_M,Y^i]
\ .
\end{align}

Finally, the symbol $\iota_Y$ denotes the interior product by a vector $Y^i$, \emph{e.g.},
\begin{align}
\iota_Y\iota_Y\left(\frac12 C_{ij}dy^i\wedge dy^j\right)=-\frac12 C_{ij}[Y^i,Y^j]
\ .
\end{align}

The bosonic effective action of an O$p^{(\epsilon_1,\epsilon_2)}$-plane in type II string theory, with $\epsilon_1,\epsilon_2\in\{-1,+1\}$, is
\be
S_{\textrm{O}p^{(\epsilon_1,\epsilon_2)}} \ 
= 
\ S_{\textrm{O}p^{(\epsilon_1,\epsilon_2)}}^{\textrm{DBI}} \, + \, S_{\textrm{O}p^{(\epsilon_1,\epsilon_2)}}^{\textrm{WZ}} \ ,
\ee
where the DBI and WZ terms are
\begin{align}
S_{\textrm{O}p^{(\epsilon_1,\epsilon_2)}}^{\textrm{DBI}}
=&\
-T_{\textrm{O}p}
\int d^{p+1}x \, e^{-\Phi}\sqrt{-\det(G_{MN})}
\ ,
\\
S_{\textrm{O}p^{(\epsilon_1,\epsilon_2)}}^{\textrm{WZ}}
=&\
\mu_{\textrm{O}p} \int\limits_{\textrm{WV}(\textrm{O}p)} C_{(p+1)} + \ldots
\ ,
\end{align}
with 
\begin{align}
\label{eq:Optension/charge}
T_{\textrm{O}p}=\epsilon_1 2^{p-4} T_{\textrm{D}p}
\ ,
&&
\mu_{\textrm{O}p}=\epsilon_2 2^{p-4} \mu_{\textrm{D}p}
\ .
\end{align}
Let us note that the theories obtained combining a stack of parallel D$p$-branes and a O$p^{(\epsilon_1, \epsilon_2)}$-plane are not supersymmetric for arbitrary values of $\epsilon_1$ and $\epsilon_2$. An intuitive way of seeing this is by noticing that the net force between these objects will not vanish (as expected for a supersymmetric configuration) unless $\epsilon_1=\epsilon_2$. Therefore, in the rest of the paper we always assume\footnote{We have nevertheless checked that this condition is also forced by the matching with supergravity.}
\begin{equation}
\epsilon_1=\epsilon_2\equiv \epsilon_{{\rm{O}}p}\,,
\label{eq:epsilonOp}
\end{equation}
and simply use the notation O$p^{\pm}$ to refer to the O$p^{(\pm, \pm)}$ planes which are BPS with respect to the D$p$-branes. In addition to this and for the sake of concreteness, we will only discuss the case $\epsilon_{\Dp p}=+1$, corresponding to D$p$-branes of possitive charge.
 
When dimensionally reducing upon the entire transverse space, the above action reduces, in its low energy limit, to a maximal SYM theory with gauge groups of C \& D type. This is due to the presence of O-planes parallel to the D-branes in our concrete setup.

\subsection{Case O5/D5 \& Open Strings}
\label{app:O5D5}

Let us compute the contribution of the sources to the scalar potential. Focusing on the scalar sector and making use of eqs.~\eqref{eq:D5Mmunu} and \eqref{eq:D5Qmn}, we find that the reduction of the DBI action of $N_{\rm D5}$ coincident D5-branes is given by 
\begin{equation}
\begin{aligned}
S_{\textrm{D}5}^{\textrm{DBI}}
=\,&
-N_{\rm D5}\,T_{\rm{D}5}\int d^{6}x\sqrt{-g}\left[\rho^{-1}\tau^{-4}\left(1+\frac{\lambda^2}{6}g_{IJK}\epsilon_{mnpq}h^{q}Y^{Im}Y^{Jn}Y^{Kp}\right)\right.\\[1mm]
&\left.+\frac{\lambda^2}{4}\rho\tau^{-4} g_{IJ}{}^{M}g_{KLM}M_{mn}M_{pq}Y^{Im}Y^{Jp}Y^{Kn}Y^{Lq}+\dots\right]
\, ,
\end{aligned}
\end{equation}
where the dots mean subleading contributions in $\lambda$ (or, equivalently, in $\alpha'$) which cannot be captured by the gauged-supergravity description together with other terms that will not enter the scalar potential.  On the other hand, for the O5 planes we get simply the contribution from the tension
\begin{equation}
S_{{\textrm{O}5}}^{\textrm{DBI}}
=\,-N_{\rm D5}\,T_{\rm{O}5}\int d^{6}x\sqrt{-g}\rho^{-1}\tau^{-4}
+\ldots\ .
\end{equation}
Hence, the total contribution from the DBI action of the sources amounts to 
\begin{equation}
\begin{aligned}
V^{\rm{DBI}}_{\rm{D}5/\rm{O}5}=\,&\rho^{-1}\tau^{-4}\left[2\kappa_6^2\left(N_{\rm D5}\,T_{\Dp 5}+T_{\rm{O}5}\right)+\frac{2\kappa_6^2\lambda^2N_{\rm D5}\,T_{\rm D5}}{6}g_{IJK}\epsilon_{mnpq}h^{q}Y^{Im}Y^{Jn}Y^{Kp}\right]\\[1mm]
&+\rho\tau^{-4}\left(\frac{2\kappa_6^2\lambda^2N_{\rm D5}\,T_{\rm D5}}{4} g_{IJ}{}^{M}g_{KLM}M_{mn}M_{pq}Y^{Im}Y^{Jp}Y^{Kn}Y^{Lq}\right)\,.
\end{aligned}
\end{equation}
This is not the only contribution of the sources to the scalar potential, as the Wess-Zumino terms in the D5-brane action give additional contributions. These, however, have been already included through the modification of the field strength \eqref{eq:D5O5Fmodified}. Let us discuss this aspect in more detail. The Wess-Zumino terms in the D5-brane action include a coupling to $C_{(8)}$, 
\begin{equation}\label{eq:couplingtoC8}
S^{\rm{WZ}}_{\rm{D}5}
=
\mu_{\Dp 5} \int_{\rm{WV(D5)}} \mathrm{Tr}\left(\mathrm{P}\left[e^{i\lambda\iota_Y\iota_Y}{\hat{C}_{(8)}}\wedge e^{\hat{B}_{(2)}}\wedge e^{\lambda\mathcal{F}}\right]\right)+\ldots \,,
\end{equation}
which modifies the Bianchi identity of $C_{(0)}$ and consequently the form of the associated field strength. Locally, we now have $F_{(1)}=dC_{(0)}+\chi_{(1)}$ for some 1-form $\chi_{(1)}$. The effect of this in our setup is that now the $F_{(1)}$-flux is no longer constant,
\begin{equation}
F_{m}=f_{m}+\Delta f_{m}\, ,
\end{equation}
since $\Delta f_{m}$ is a certain combination of the non-Abelian scalars $Y^{Im}$. This will result in two additional contributions to the scalar potential coming from the kinetic term of $F_{(1)}$, namely:
\begin{equation}\label{eq:VF1}
V_{F_{(1)}}=\frac{\rho\,\tau^{-6}}{2}M^{mn}\left(f_m+\Delta f_m\right)\left(f_n+\Delta f_n\right)\, .
\end{equation}
By a standard argument, the term in the potential which is linear in $\Delta f_{m}$ can be read by evaluating \eqref{eq:couplingtoC8} using the uncorrected expression for ${\hat C}_{(8)}$.\footnote{Namely, the one obtained solving $d C_{(8)}=\star d C_{(0)}$, with $C_{(0)}$ given by \eqref{eq:O5C0}.} Let us do this explicitly in order to show how one can get from a direct calculation the expression of the modified field strength presented in the main text, \eqref{eq:D5O5Fmodified}. 

First, we expand the integrand of \eqref{eq:couplingtoC8} at the relevant order in $\lambda$:
\begin{equation}\label{eq:couplingtoC82}
\begin{aligned}
 &\mu_{\Dp 5} \int \mathrm{Tr}\left(\mathrm{P}\left[e^{i\lambda\iota_Y\iota_Y}{\hat{C}_{(8)}}\wedge e^{\hat{B}_{(2)}}\wedge e^{\lambda\mathcal{F}}\right]\right)
=\,i\lambda \mu_{\Dp 5} \int{\rm{Tr}}\, \iota_{Y}\iota_{Y}{\hat C}_{(8)}+\dots\\[1mm]
=&\,-\frac{i\lambda \mu_{\Dp 5}}{2! 6!} \int dx^{\mu_1}\wedge \dots \wedge dx^{\mu_6}\, {\rm{Tr}} \left({\hat C}_{(8)}{}_{\mu_1\dots\mu_6 mn}[Y^m, Y^n]\right)+\dots\\[1mm]
\end{aligned}
\end{equation}
Now we insert the expression of ${\hat C}_{(8)}$, which is the following\footnote{Our conventions are such that $\epsilon_{01 \dots d-1}=+\sqrt{-g^{(d)}}$.}
\begin{equation}
{{\hat C}_{(8)}{}_{\mu_1 \dots \mu_6 mn}=\frac{\lambda}{3}\,\epsilon_{\mu_1\dots \mu_6 mnp}{}^{q}f_q Y^p\,.}
\end{equation}
Plugging this in \eqref{eq:couplingtoC82} yields\footnote{$\epsilon_{1234}=+\sqrt{{\rm{det}}M_{mn}}=+1$.}
\begin{equation}
\begin{aligned}
&\mu_{\Dp 5}\int \mathrm{Tr}\left(\mathrm{P}\left[e^{i\lambda\iota_Y\iota_Y}{\hat{C}_{(8)}}\wedge e^{\hat{B}_{(2)}}\wedge e^{\lambda\mathcal{F}}\right]\right)
=\\[1mm]
=\,&\frac{\lambda^2 N_{\rm D5}\, \mu_{\Dp 5}}{3!}\int d^6x \sqrt{-g}\rho \tau^{-6}\,g_{IJK}\epsilon_{mnpq}M^{qr}f_r Y^{Im}Y^{Jn}Y^{Kp}+\dots\,.
\end{aligned}
\end{equation}
Comparing this with the term in the scalar potential \eqref{eq:VF1} which is linear in $\Delta f_{m}$, we obtain
\begin{equation}
\Delta f_{m}=-\frac{2\kappa_6^2\lambda^2N_{\rm D5}\,\mu_{\Dp 5}}{3!}g_{IJK}\epsilon_{mnpq}Y^{In}Y^{Jp}Y^{Kq}\, ,
\end{equation}
as anticipated in \eqref{eq:D5O5Fmodified}.

%%%%%%%%%%%%%%%%%%%%%%%%%%%%%%%%%%%%%%%%%%%%%%%%%%%%%%%%%%%%%%%%%%%%%%%%%%%%%%%%%%%%%%%%%%%%%%%%%%%%%%%%%%%%%%%%%%%%%

\subsection{Case O7/D7 \& Open Strings}
\label{app:O7D7}

In this section we will explain how to obtain the modified field strengths $F_{(1)}$ and $F_{(3)}$ in \eqref{eq:D7F1mod} and \eqref{eq:D7F3mod} from the WZ effective actions. We will study the case of $F_{(1)}$ and give similar arguments for $F_{(3)}$.

Let us firstly consider the contribution of the $C_{(8)}$ potential in the bulk and the WZ actions.\footnote{Note that the DBI action does not contribute to the equation of motion of the RR fields.}

Using the type IIB democratic formulation  \eqref{IIB_action}, together with the duality relation $F_{(9)}=\star F_{(1)}$, the variation of 
\begin{align}
S=S_{\text{IIB}}+S^{\text{WZ}}_{\text{D}7}+S^{\text{WZ}}_{\text{O}7}
\end{align}
with respect to $C_{(8)}$ is
\begin{align}
\delta S_{\text{total}}
=
\int_{10} \left(
	\frac{1}{2\kappa_{10}^2}\, d\star F_{(9)}
	+(\star J)_{(2)}
	\right)\wedge \delta C_{(8)}
\ ,
\end{align}
where we have rewritten the WZ action as
\begin{align}
S^{\text{WZ}}_{\text{O}p/\text{D}p}=\int_{\text{WV}(\text{O}p/\text{D}p)} \omega^{(p+1)}
=\int_{10} \omega^{(p+1)}\wedge \delta^{\text{O}p/\text{D}p}_{9-p}
\ ,
\end{align}
with $\delta^{\text{D}p}_{9-p}\equiv \delta(x^{p+1})\cdots \delta(x^{p+1})\, dx^{p+1}\wedge \cdots \wedge dx^9$.  The quantity  $(\star J)_{(2)}$ is defined through the WZ action as follows:
\begin{align}
S_{\text{D}7}^{\text{WZ}}
=
\int_{10} C_{(8)}\wedge (\star J)_{(2)}
\ .
\label{eq:D7C8J2}
\end{align}
Therefore, the $C_{(8)}$ equation of motion becomes the modified Bianchi identity for $F_{(1)}$,
\begin{align}
 d F_{(1)}=
	-\, 2\, \kappa_{8}^2\, (\star J)_{(2)}
\ ,
\label{eq:starJ2}
\end{align}
in such a way the source term is straightforwardly determined from the $C_{(8)}$ couplings in the expansion of the WZ action. On the other hand, taking into account that the product $\sigma_{\text{O}7} (-1)^{F_L} \Omega$ on the components $(C_{\mu_0\cdots \mu_5 ab}, C_{\mu_0\cdots \mu_5ai},C_{\mu_0\cdots \mu_5ij}) $ is, respectively, $(+,-,+)$, only the quantities $(\star J)_{ab}$ and $(\star J)_{ij}$ will be nonzero. Here, because the flux $\overline{F}_a$ is allowed while $\overline{F}_i$ is projected out, we will focus on $(\star J)_{ab}$.

In this case, we have to consider the following $C_{(8)}$ couplings:
\begin{align}
S_{\text{D}7}^{\text{WZ}}
=&\
\mu_{\text{D}7}\int \Tr{
	\text{P}[ \hat C_{(8)}] 
	+i\lambda^2\, \text{P}[\iota_Y\, \iota_Y\, \hat{C}_{(8)}]\wedge \cF
	\phantom{\frac{1}{2}}
	\right.
\nonumber\\
&\
\left.
\qquad\qquad\qquad
	-\frac12\lambda^2\,\text{P}[(\iota_Y\, \iota_Y)^2\,\hat  C_{(8)}\wedge \hat B\wedge \hat B]
	}
+\ldots
\ ,
\label{eq:D7WZ}
\\
S_{\text{O}7}^{\text{WZ}}
=&\
\mu_{\text{O}7}\int C_{(8)}
+\ldots
\ .
\end{align}
The first contribution can be conveniently rewritten as
\begin{align}
\int \Tr{\text{P}[ \hat C_{(8)}] }
=
\int \frac{1}{6!2!}\,\Tr{\text{P}[ \hat C_{(8)}]_{\mu_0\cdots\mu_5\underline{ab}} }\,\delta(y^1)\,\delta(y^2)\,dx^{\mu_0\cdots\mu_5\underline{ab}}\wedge \frac{\epsilon_{\underline{ij}}}{2}\,dy^{\underline{ij}}
\ ,
\end{align}
where we have used the notation $dx^{M_1\cdots M_n}\equiv dx^{M_1}\wedge\cdots\wedge dx^{M_n}$.

The pullback of the 8-form potential is expressed as
\begin{align}
\text{P}[\hat C_{(8)}]_{\mu_0\cdots\mu_5\underline{ab}}
=&\
\hat C_{\mu_0\cdots\mu_5\underline{ab}}
-\lambda D_{[\mu_0}Y^{\underline{k}}\, \hat C_{\mu_1\cdots\mu_5\underline{ab}]\underline{k}}\,
%\\
%&\
+\frac{\lambda^2}{2} D_{[\mu_0}Y^{\underline{k}}\, D_{\mu_1}Y^{\underline{l}}\,\hat C_{\mu_2\cdots\mu_5\underline{ab}]\underline{kl}}
+\ldots
\ ,
\label{eq:pullbackC8}
\end{align}
where, moreover, $\hat C_{(8)}$ is Taylor-expanded around the position of the source, $y^i=0$,
\begin{align}
\hat C_{\mu_0\cdots\mu_5ab}
=&\
C_{\mu_0\cdots\mu_5ab}
+\lambda\,  Y^{\underline{i}}\, \partial_{\underline{i}} \, C_{\mu_0\cdots\mu_5\underline{ab}}
+\frac{\lambda^2}{2}\, Y^{\underline{i}}\, Y^{\underline{j}} \, \partial_{\underline{i}}\, \partial_{\underline{j}} \, C_{\mu_0\cdots\mu_5\underline{ab}}
+\ldots
\ .
\label{eq:hatC8}
\end{align}
Here, $C_{\mu_0\cdots\mu_5\underline{ab}}$ is a 10D field, which admits the Kaluza Klein decomposition
\begin{align}
C_{\mu_0\cdots\mu_5\underline{ab}}(x,y)
=
C_{\mu_0\cdots\mu_5mn}(x) \, v^m{}_{\underline{a}}(y)\, v^n{}_{\underline{b}}(y)
+\ldots
\ ,
\end{align}
where ellipses account for other nontrivial terms entering the compactification Ansatz that do not depend on the 8-form potential which will be omitted in this analysis without loss of generality.

All in all, we find that the first term in \eqref{eq:pullbackC8} gives the following contribution to $(\star J)_{ab}$:
\begin{align}
\hat C_{\mu_0\cdots\mu_5\underline{ab}}
=
v^a{}_{\underline{a}} \, v^b{}_{\underline{b}}\,  \frac{\lambda^2}{2}\, Y^{Ii'}\, Y^{Jj'}\, \omega_{ai'}{}^k \, \omega_{bj'}{}^l \, C_{\mu_0\cdots\mu_5kl}\, t_I\, t_J
\ .
\end{align}
If we multiply both sides of the equation by $\epsilon_{\underline{ij}}$ and use the Schouten identity on the $\underline{kl}$ and $\underline{ij}$ indices, we have
\begin{align}
\hat C_{\mu_0\cdots\mu_5\underline{ab}}\epsilon_{ij}
=
v^a{}_{\underline{a}} \, v^b{}_{\underline{b}}\, \frac{\lambda^2}{2}\, Y^{Ii'}\, Y^{Jj'}\, \omega_{ai'}{}^k \, \omega_{bj'}{}^l \, C_{\mu_0\cdots\mu_5ij}\epsilon_{kl}\, t_I\, t_J
\ .
\end{align}

Let us consider now the second term of \eqref{eq:pullbackC8}. Using the twist matrices \eqref{eq:D7twist}, this term contains the quantity
\begin{multline}
-\lambda D_{[\mu_0}Y^{\underline{k}}\, \hat C_{\mu_1\cdots\mu_5\underline{ab}]\underline{k}}
=
-v^a{}_{\underline{a}} \, v^b{}_{\underline{b}}\, \frac{8\cdot 7}{2}\frac{\lambda^2}{2} \, g_{JK}{}^I\, 
\\
\times \cA^J{}_{[a|}\, Y^{Kk} Y^{Li} \left(
	\omega_{i|b]}{}^j\, C_{\mu_0\cdots\mu_5jk}
	+\omega_{ik}{}^c\, C_{\mu_0\cdots\mu_5|b]c}
	\right) t_I\, t_L
+\ldots
\ .
\end{multline}
Then, while the first term contributes to $(\star J)_{ab}$ (we need to use the Schouten identity on the indices $[\underline i\underline j\underline k]$ and take the trace), the latter does not, due to the presence of the longitudinal indices $[bc]$ in the potential. In particular, the full term is given by
\begin{align}
\Tr{-\lambda D_{[\mu_0}Y^{\underline{k}}\, \hat C_{\mu_1\cdots\mu_5\underline{ab}]\underline{k}}}
=
 v^a{}_{\underline{a}} \, v^b{}_{\underline{b}}\, \frac{8\cdot 7}{4}\, N_{\rm D7}\,\frac{\lambda^2}{2} \, g_{JKL}\, 
 \eta_{[a}\,
 \cA^J{}_{b]}\, Y^{Kk} Y^{Li}  \, C_{\mu_0\cdots\mu_5ki}
+\ldots
\ .
\end{align}

The last term in  \eqref{eq:pullbackC8} can be written as
\begin{multline}
\frac{\lambda^2}{2} D_{[\mu_0}Y^{\underline{k}}\, D_{\mu_1}Y^{\underline{l}}\,\hat C_{\mu_2\cdots\mu_5\underline{ab}]\underline{kl}}
=
v^a{}_{\underline{a}} \, v^b{}_{\underline{b}}\, \frac{8\cdot 7}{2}\,
\frac{\lambda^2}{2} \cA^J{}_{[a|} \, \cA^{J'}{}_{|b]}\, Y^{Kk} \, Y^{K'l} \, 
\\
\times g_{JK}{}^I \, g_{J'K'}{}^{I'} \, C_{\mu_0\cdots\mu_5kl}\, t_I\, t_{I'}
+\ldots
\ .
\end{multline}
Multiplying by $\epsilon_{\underline i \underline j}$ and using again the Schouten identity, we have
\begin{multline}
\frac{\lambda^2}{2} D_{[\mu_0}Y^{\underline{k}}\, D_{\mu_1}Y^{\underline{l}}\,\hat C_{\mu_2\cdots\mu_5\underline{ab}]\underline{kl}}\,\epsilon_{ij}
\\
=
v^a{}_{\underline{a}} \, v^b{}_{\underline{b}}\,
\frac{8\cdot 7}{2}\,
\frac{\lambda^2}{2} \cA^J{}_{[a|} \, \cA^{J'}{}_{|b]}\, Y^{Kk} \, Y^{K'l} \, g_{JK}{}^I \, g_{J'K'}{}^{I'} \, C_{\mu_0\cdots\mu_5ij}\, \epsilon_{kl}\, t_I\, t_{I'}
+\ldots
\ ,
\label{eq:DYDY-WZ-D7}
\end{multline}
where we have omitted terms proportional to $D_\mu$.

Next, by studying the second term coming from the WZ action \eqref{eq:D7WZ}, we observe that the only contribution to $(\star J)_{ab}$ arises from
\begin{multline}
\Tr{i\lambda^2\, \text{P}[\iota_Y\, \iota_Y\, \hat{C}_{(8)}]\wedge \cF}
=
-N_{\rm D7}\, \frac{\lambda^2}{2}\, \kappa_{I I'}\,  g_{JK}{}^I \, g_{J'K'}{}^{I'} 
\\
\times \, \cA^{I}{}_a \, \cA^{J}{}_b \,Y^{I'i}\, Y^{J'j}\, \epsilon_{ij}\, C_{\mu_0\cdots\mu_5 k l} \frac{dx^{\mu_0\cdots\mu_5}}{6!}\wedge \frac{v^{kl}}{2!}\wedge \frac{v^{ab}}{2!}
+\ldots
\ ,
\end{multline}
where we have used the notation $v^{mn}\equiv v^m\wedge v^n$.
Precisely, this term cancels the one in \eqref{eq:DYDY-WZ-D7}. 

Finally, let us consider the third and last term in the WZ action \eqref{eq:D7WZ},  $\text{P}[(\iota_Y\, \iota_Y)^2\,\hat  C_{(8)}]$. This consists of the 4th interior product over the vector $Y^{\underline{i}}$ of a 12-form. Because we are dealing with codimension-2 objects, this turns out to be trivially zero. 

Therefore, according to \eqref{eq:D7C8J2}, the final expression for $(\star J)_{ab}$ is
\begin{align}
(\star J)_{ab}
=
 N_{\rm D7}\, \mu_{\text{D}7}\, \frac{ \lambda^2}{2}\left(
 	\kappa_{II'} \,\epsilon_{kl} 
	Y^{Ii'}\, Y^{I'j'}\, \omega_{ai'}{}^k \, \omega_{bj'}{}^l 
	+  g_{JKL}\, 
 \eta_{[a}\,
 \cA^J{}_{b]}\, Y^{Kk} Y^{Li}  \epsilon_{ki}
	\right)
	\ ,
\end{align}
which implies, using \eqref{eq:starJ2},
\begin{multline}
dF_{(1)}
=
-2\, \kappa_{8}^2\, N_{\rm D7}\, \mu_{\text{D}7}\,  \frac{ \lambda^2}{2\cdot 2!}\left(
 	\kappa_{II'} \,\epsilon_{kl} 
	Y^{Ii'}\, Y^{I'j'}\, \omega_{ai'}{}^k \, \omega_{bj'}{}^l 
	\right.
\\	
\left.
+  g_{JKL}\, 
 \eta_{[a}\,
 \cA^J{}_{b]}\, Y^{Kk} Y^{Li}  \epsilon_{ki}
	\right)\, v^a\wedge v^b
+\ldots
\ .	
\label{eq:dF1WZ}
\end{multline}
Then, because generically its internal part is $F_{(1)} = F_a\, v^a+ F_i\, v^i$ and $\omega_{ab}{}^i=0$, the only contribution to $(\star J)_{ab}$ arises from $F_a$. In particular,
\begin{align}
 d F_{(1)}= -\frac12 F_{a}\, (\omega_{ij}{}^a \, v^i\wedge v^j +\, \omega_{bc}{}^a \,v^b\wedge v^c)
 +\ldots
\ .
\end{align}
Equating the $v^a\wedge v^b$ components with \eqref{eq:dF1WZ} and using the Jacobi identities \eqref{eq:Jacobi-omega}, in particular the first equation in \eqref{eq:QCD7O7}, we obtain
\begin{align}
F_a= \frac{\tilde\lambda_7^2}{2}\left(
	 \kappa_{ai}{}^j\, \kappa_{IJ}\, Y^{Ii}\, Y^{Jk} \epsilon_{jk} 
	 -  g_{JKL}\,  \cA^J{}_{a}\, Y^{Kk} Y^{Li}  \epsilon_{ki}
	 \right)
	 +\Delta_a
\ ,
\qquad
\epsilon^{ab}\, \Delta_a\, \eta_b=0 
\ ,
\end{align}
where $\tilde \lambda_7\equiv (2\kappa_{8}^2\, T_{\text{D}7})^{1/2}\lambda$. Precisely, the compactification Ansatz for $ C_{(0)}$, together with the integrability condition \eqref{eq:integrabilityF1} allow us to identify $\Delta_a=\overline{F}_a$, in such a way that \eqref{eq:D7F1mod} is recovered.

A similar argument applies to the field strength $F_{(3)}$  in \eqref{eq:D7F3mod}. In this case we need to study the couplings to $C_{(6)}$ in the WZ action, so that we can read off the current $(\star J)_{abij}$, which is defined via this expression:
\begin{align}
S^{\text{WZ}}_{\text{D}7}
=\int_{10} C_{(6)}\wedge (\star J)_{(4)}
\ .
\end{align}
Namely, as we are interested in the couplings to $C_{\mu_0\cdots\mu_5}$, from the WZ action we have the following contributions:
\begin{multline}
S^{\text{WZ}}_{\text{D}7}
=
\mu_{\text{D}7}
\int\Tr{
	\text{P}[\hat C_{(6)}\wedge\hat B_{(2)}]
	+\lambda\,\text{P}[\hat C_{(6)}]\wedge\cF
	+\frac{i}{2}\lambda\, \text{P}[\iota_Y\iota_Y(\hat C_{(6)} \wedge \hat B_{(2)}^2)]
	\right.
\\
\left.
	+\frac{i}{2}\lambda^2\, \text{P}[\iota_Y\iota_Y(\hat C_{(6)}\wedge \hat B_{(2)})]\wedge\cF
	-\frac12\lambda^2\, \text{P}[(\iota_Y\iota_Y)^2(\hat C_{(6)}\wedge \hat B_{(2)}^3)]
	}
+\ldots
\ .
\end{multline}
The two terms in the second line do not contribute to the current: 
While the latter is trivially zero because the D7 has codimension 2,  the former is $\cO(\lambda^2)$ and the scalars $B_{ij}$ are projected out by the O7 plane.  Then, according to \eqref{eq:D7F3mod} and the compactification Ansatz for $\hat B_{(2)}$, it is expected that only the first and second terms give nontrivial contributions.

%%%%%%%%%%%%%%%%%%%%%%%%%%%%%%%%%%%%%%%%%%%%%%%%%%%%%%%%%%%%%%%%%%%%%%%%%%%%%%%%%%%%%%%%%%%%%%%%%%%%%%%%%%%%%%%%%%%%%

\subsection{Case O9/D9 \& Open Strings}
\label{app:O9D9}

Let us firstly consider the D9/O9 contributions to the scalar potential. The two-derivative action of a stack of $N_{\Dp9}$ D9-branes is given by $S_{\text{D}9}=S_{\textrm{D}9}^{\textrm{DBI}} + S_{\textrm{D}9}^{\textrm{WZ}}$, where
\begin{align}
S_{\textrm{D}9}^{\textrm{DBI}} 
\, 
= &\ 
-N_{\Dp9} \, T_{\textrm{D}9}\int{d^{10}x\, \sqrt{-G}\, e^{-{\Phi}}\left(
	1+\frac{\lambda^2}{4}\, {\cal F}^I{}_{\mu\nu}  {\cal F}^{I\, \mu\nu}
	\right)}  
\,+\,\ldots\ ,
\\
S_{\textrm{D}9}^{\textrm{WZ}} \, 
=&\ 
\mu_{\text{D}9}\int 
\mathrm{Tr}\left(
C_{(10)}
+ \frac{\lambda^2}{2} \, C_{(6)}\wedge \, \cF\wedge \cF\right)
\,+\, \ldots
\ .
\end{align}
On the other hand, the O9-plane contribution is
\begin{align}
S_{\textrm{O}9}
\, 
= &\ 
- \, T_{\textrm{O}9}\int d^{10}x\, \sqrt{-G}\, e^{-{\Phi}}
\,+\,\mu_{\text{O}9} \int C_{(10)}
\,+\,\ldots
\ ,
\end{align}
where the O9-plane charge is $\mu_{\text{O}9}= 32 \,\epsilon_{{\rm{O}}9} \, \mu_{\text{D}9}$. 
The tadpole cancellation condition,
\begin{equation}
N_{\Dp9}\mu_{{\rm D}9}+\mu_{{\rm O}9}=0\,,
\end{equation}
requires $N_{\rm D9}=32$  and $\epsilon_{{\rm O}9}=-1$, which corresponds to an $\textrm{O}9^{-}$ plane. Hence, the total contribution from the sources, $S_{{\rm D}9/{\rm O}9}=S_{{\rm D}9}+S_{{\rm O}9}$, amounts to 
\begin{equation}
S_{{\rm D}9/{\rm O}9}=-\lambda^2 N_{\rm D9} T_{\Dp 9}\int d^{10}x \sqrt{-G} \left[\frac{e^{-\Phi}}{4}\,{\cal F}^I_{\mu\nu}{\cal F}^{I}{}^{\mu\nu}+\frac{1}{2\cdot 6! \cdot (2 !)^2}\, \epsilon^{\mu_1 \dots \mu_{10} } C_{(6)}{}_{\mu_1 \dots \mu_6}{\cal F}^{I}{}_{\mu_7 \mu_8}{\cal F}_{I}{}_{\mu_9\mu_{10}}\right]\,,
\end{equation}
where adjoint indices are lowered using the Cartan-Killing metric, ${\cal F}_{I}=\kappa_{IJ}{\cal F}^{J}$. Taking into account the supergravity fields that survive the O9 projection (see Table~\ref{Table:O9_Projection}), we can write down the full action as
\begin{equation}\label{eq:typeIaction}
\begin{aligned}
S=\,&\frac{1}{2\kappa^2_{10}}\int d^{10}x \sqrt{-G}\left\{e^{-2\Phi}\left[{\cal R}+4(\partial\Phi)^2\right]-\frac{1}{2\cdot 7!}\left|F_{(7)}\right|^{2}-\frac{{\tilde \lambda}^2_9}{4}e^{-\Phi}{\cal F}^I_{\mu\nu}{\cal F}^{I}{}^{\mu\nu}\right.\\[1mm]
 &\left.- \frac{{\tilde \lambda}^2_9}{2 \cdot 6! \cdot (2!)^2}\epsilon^{\mu_1 \dots \mu_{10} } C_{(6)}{}_{\mu_1 \dots \mu_6}{\cal F}^{I}{}_{\mu_7 \mu_8}{\cal F}_{I}{}_{\mu_9\mu_{10}}\right\}\,,
\end{aligned}
\end{equation}
where 
\begin{equation}
{\tilde\lambda}^2_9\equiv 2\kappa^2_{10}\lambda^2 N_{\rm D9}\,T_{\rm {D9}}\, .
\end{equation}
This is nothing but the bosonic effective action of type I string theory, written in terms of a RR 6-form potential $C_{(6)}$. In order to write it down in the standard form, we dualize it into a 2-form potential $C_{(2)}$. To this aim, we integrate by parts the last term in \eqref{eq:typeIaction} and introduce a Lagrange multiplier, as usual. Using differential-form notation, we have
\begin{equation}
\begin{aligned}
S\to S'
=&\,
S
-\frac{{\tilde\lambda}^2_9}{4\kappa^2_{10}}\int d\left(C_{(6)}\wedge \Omega_{(3)}\right)
+\frac{1}{2\kappa^2_{10}}\int F_{(7)}\wedge d C_{(2)}
\\[1mm]
=&\, 
\frac{1}{2\kappa^2_{10}}\int \left\{\frac{1}{2} F_{(7)}\wedge \star F_{(7)}
+F_{(7)}\wedge \left(dC_{(2)}-\frac{{\tilde\lambda}^2_9}{2}\,\Omega_{(3)}\right)\right\}
+\ldots
\\[1mm]
=&\,
\frac{1}{2\kappa^2_{10}}\int \left\{\frac{1}{2} F_{(7)}\wedge \star F_{(7)}+F_{(7)}\wedge F_{(3)}\right\}\,,
\end{aligned}
\end{equation}
where we have defined the Chern-Simons 3-form $\Omega_{(3)}$,
\begin{equation}
\Omega_{(3)}=\cF^I\wedge \cA^I+\frac{1}{3!}\,g_{IJK}\, \cA^I\wedge \cA^J\wedge \cA^K\,, \hspace{1cm} d\Omega_{(3)}=\cF^I\wedge \cF_{I}\,, 
\end{equation}
as well as the modified field strength $F_{(3)}$,
\begin{equation}
\begin{aligned}
F_{(3)}=\,&dC_{(2)}-\frac{{\tilde\lambda}^2_9}{2}\,\Omega_{(3)}\\[1mm]
=\,&dC_{(2)}-\frac{{\tilde\lambda}^2_9}{2}\left(\cF^I\wedge \cA^I+\frac{1}{3!}\,g_{IJK}\, \cA^I\wedge \cA^J\wedge \cA^K\right)\, .
\end{aligned}
\end{equation}
The variation of $S'$ with respect to $F_{(7)}$ (now considered non-dynamical) gives 
\begin{equation}\label{eq:dualityrelation}
F_{(3)}=-\star F_{(7)}\, .
\end{equation}
whereas the variation with respect to $C_{(2)}$ gives the Bianchi identity of $F_{(7)}$, namely $dF_{(7)}=0$. The Bianchi identity of $F_{(3)}$ is now modified as a consequence of the coupling of the open string-sector to $C_{(6)}$. It reads
\begin{equation}\label{eq:BianchiF3}
dF_{(3)}=-\frac{{\tilde\lambda}^2_9}{2} \cF^{I}\wedge \cF_{I}\, .
\end{equation}
Finally, we substitute the duality relation \eqref{eq:dualityrelation} back into the action. This yields the bosonic action of type I supergravity, as anticipated:
\begin{equation}
S'=\frac{1}{2\kappa^2_{10}}\int d^{10}x\sqrt{-G}\left\{e^{-2\Phi}\left[{\cal R}+4(\partial\Phi)^2\right]-\frac{1}{2\cdot 3!}\left|F_{(3)}\right|^{2}-\frac{{\tilde\lambda}^2_9}{4}e^{-\Phi}{\cal F}^I_{\mu\nu}{\cal F}^{I}{}^{\mu\nu}\right\}
\end{equation}

When considering the reduction Ansatz \eqref{eq:D9-ansatz} and \eqref{eq:D9-ansatz2}, the internal components of the field strength ${\cal F}^{I}$ are
\begin{align}
\cF^I
=&\
\frac12\left(
	\overline\cF^I{}_{mn}
	-\, g_{JK}{}^I \, \cA^J{}_m \, \cA^K{}_n
	-\cA^I{}_p \, \omega_{mn}{}^p
	\right)v^m \wedge v^n
	+\ldots 
\ ,
\end{align}
where we have assumed that 
\begin{align}
g_{IJK}\, \sigma^I = 0
\ ,
\qquad\qquad 
d\sigma^I = \frac12 \, \overline\cF^I{}_{mn} \,v^m \wedge v^n
\ ,
\end{align}
for constant $\overline\cF^I{}_{mn}$. The latter imposes the integrability condition 
\begin{align}
\overline\cF^{I}{}_{q[m}\, \omega_{np]}{}^q
=
0
\ ,
\end{align}
which turns out to be a quadratic constraint in supergravity \eqref{eq:D9-QC}.

Similarly, the internal components of $F_{(3)}$ result
\begin{multline}
F_{mnp}
=
	\overline F_{mnp}-3 \,C_{q[m}\,\omega_{np]}{}^q 
\\	-\,\tilde\lambda_9^2\left(
		3\,\cA^I{}_m \, \overline\cF^I{}_{np}
		-\, g_{IJK}\, \cA^I{}_m \,\cA^J{}_n \,\cA^K{}_p
		-\frac32\, \omega_{mn}{}^q \,\cA^I{}_p \,\cA^I{}_q
		\right)
\ ,
\end{multline}
where we have introduced
\begin{align}
d\gamma-\frac{\tilde\lambda_9^2}{2}
\sigma^I\wedge d\sigma^I
\equiv
\frac{1}{3!}\, \overline F_{mnp} v^m\wedge v^n \wedge v^p
\ ,
\end{align}
for constant ${\overline F}_{mnp}$. This leads to the integrability condition
\begin{align}\label{eq:intcondgamma}
\frac{\tilde \lambda_9^2}{2}\, \overline \cF^I{}_{[mn} \, \overline\cF^I_{pq]}
-\overline F_{r[mn}\,\omega_{pq]}{}^r
=0
\ ,
\end{align}
which is again a quadratic constraint, \eqref{eq:D9-QC}.

\newpage

\bibliographystyle{utphys}
\bibliography{IIB_Orientifolds_Open_arxiv}

\providecommand{\href}[2]{#2}\begingroup\raggedright\begin{thebibliography}{10}

\bibitem{Vafa:2005ui}
C.~Vafa, ``{The String landscape and the swampland},''
  \href{http://arxiv.org/abs/hep-th/0509212}{{\ttfamily arXiv:hep-th/0509212}}.

\bibitem{Ooguri:2006in}
H.~Ooguri and C.~Vafa, ``{On the Geometry of the String Landscape and the
  Swampland},'' \href{http://dx.doi.org/10.1016/j.nuclphysb.2006.10.033}{{\em
  Nucl. Phys. B} {\bfseries 766} (2007) 21--33},
  \href{http://arxiv.org/abs/hep-th/0605264}{{\ttfamily arXiv:hep-th/0605264}}.

\bibitem{Adams:2010zy}
A.~Adams, O.~DeWolfe, and W.~Taylor, ``{String universality in ten
  dimensions},'' \href{http://dx.doi.org/10.1103/PhysRevLett.105.071601}{{\em
  Phys. Rev. Lett.} {\bfseries 105} (2010) 071601},
  \href{http://arxiv.org/abs/1006.1352}{{\ttfamily arXiv:1006.1352 [hep-th]}}.

\bibitem{Cvetic:2020kuw}
M.~Cveti\v{c}, M.~Dierigl, L.~Lin, and H.~Y. Zhang, ``{String Universality and
  Non-Simply-Connected Gauge Groups in 8d},''
  \href{http://dx.doi.org/10.1103/PhysRevLett.125.211602}{{\em Phys. Rev.
  Lett.} {\bfseries 125} no.~21, (2020) 211602},
  \href{http://arxiv.org/abs/2008.10605}{{\ttfamily arXiv:2008.10605
  [hep-th]}}.

\bibitem{Cvetic:2021sjm}
M.~Cvetic, M.~Dierigl, L.~Lin, and H.~Y. Zhang, ``{Gauge group topology of 8D
  Chaudhuri-Hockney-Lykken vacua},''
  \href{http://dx.doi.org/10.1103/PhysRevD.104.086018}{{\em Phys. Rev. D}
  {\bfseries 104} no.~8, (2021) 086018},
  \href{http://arxiv.org/abs/2107.04031}{{\ttfamily arXiv:2107.04031
  [hep-th]}}.

\bibitem{Bedroya:2021fbu}
A.~Bedroya, Y.~Hamada, M.~Montero, and C.~Vafa, ``{Compactness of brane moduli
  and the String Lamppost Principle in d \ensuremath{>} 6},''
  \href{http://dx.doi.org/10.1007/JHEP02(2022)082}{{\em JHEP} {\bfseries 02}
  (2022) 082}, \href{http://arxiv.org/abs/2110.10157}{{\ttfamily
  arXiv:2110.10157 [hep-th]}}.

\bibitem{Taylor:2019ots}
W.~Taylor and A.~P. Turner, ``{Generic matter representations in 6D
  supergravity theories},''
  \href{http://dx.doi.org/10.1007/JHEP05(2019)081}{{\em JHEP} {\bfseries 05}
  (2019) 081}, \href{http://arxiv.org/abs/1901.02012}{{\ttfamily
  arXiv:1901.02012 [hep-th]}}.

\bibitem{Fraiman:2022aik}
B.~Fraiman and H.~Parra De~Freitas, ``{Unifying the 6D $ \mathcal{N} $ = (1, 1)
  string landscape},'' \href{http://dx.doi.org/10.1007/JHEP02(2023)204}{{\em
  JHEP} {\bfseries 02} (2023) 204},
  \href{http://arxiv.org/abs/2209.06214}{{\ttfamily arXiv:2209.06214
  [hep-th]}}.

\bibitem{Fraiman:2018ebo}
B.~Fraiman, M.~Gra\~na, and C.~A. N\'u\~nez, ``{A new twist on heterotic string
  compactifications},'' \href{http://dx.doi.org/10.1007/JHEP09(2018)078}{{\em
  JHEP} {\bfseries 09} (2018) 078},
  \href{http://arxiv.org/abs/1805.11128}{{\ttfamily arXiv:1805.11128
  [hep-th]}}.

\bibitem{Font:2020rsk}
A.~Font, B.~Fraiman, M.~Gra\~na, C.~A. N\'u\~nez, and H.~P. De~Freitas,
  ``{Exploring the landscape of heterotic strings on $T^d$},''
  \href{http://dx.doi.org/10.1007/JHEP10(2020)194}{{\em JHEP} {\bfseries 10}
  (2020) 194}, \href{http://arxiv.org/abs/2007.10358}{{\ttfamily
  arXiv:2007.10358 [hep-th]}}.

\bibitem{Font:2021uyw}
A.~Font, B.~Fraiman, M.~Gra\~na, C.~A. N\'u\~nez, and H.~Parra De~Freitas,
  ``{Exploring the landscape of CHL strings on T$^{d}$},''
  \href{http://dx.doi.org/10.1007/JHEP08(2021)095}{{\em JHEP} {\bfseries 08}
  (2021) 095}, \href{http://arxiv.org/abs/2104.07131}{{\ttfamily
  arXiv:2104.07131 [hep-th]}}.

\bibitem{Angelantonj:2003rq}
C.~Angelantonj, S.~Ferrara, and M.~Trigiante, ``{New D = 4 gauged
  supergravities from N=4 orientifolds with fluxes},''
  \href{http://dx.doi.org/10.1088/1126-6708/2003/10/015}{{\em JHEP} {\bfseries
  10} (2003) 015}, \href{http://arxiv.org/abs/hep-th/0306185}{{\ttfamily
  arXiv:hep-th/0306185}}.

\bibitem{Angelantonj:2003up}
C.~Angelantonj, S.~Ferrara, and M.~Trigiante, ``{Unusual gauged supergravities
  from type IIA and type IIB orientifolds},''
  \href{http://dx.doi.org/10.1016/j.physletb.2003.12.055}{{\em Phys. Lett. B}
  {\bfseries 582} (2004) 263--269},
  \href{http://arxiv.org/abs/hep-th/0310136}{{\ttfamily arXiv:hep-th/0310136}}.

\bibitem{Roest:2009dq}
D.~Roest, ``{Gaugings at angles from orientifold reductions},''
  \href{http://dx.doi.org/10.1088/0264-9381/26/13/135009}{{\em Class. Quant.
  Grav.} {\bfseries 26} (2009) 135009},
  \href{http://arxiv.org/abs/0902.0479}{{\ttfamily arXiv:0902.0479 [hep-th]}}.

\bibitem{Angelantonj:2003zx}
C.~Angelantonj, R.~D'Auria, S.~Ferrara, and M.~Trigiante, ``{K3 x T**2 / Z(2)
  orientifolds with fluxes, open string moduli and critical points},''
  \href{http://dx.doi.org/10.1016/j.physletb.2003.12.074}{{\em Phys. Lett. B}
  {\bfseries 583} (2004) 331--337},
  \href{http://arxiv.org/abs/hep-th/0312019}{{\ttfamily arXiv:hep-th/0312019}}.

\bibitem{Andriot:2022bnb}
D.~Andriot, P.~Marconnet, M.~Rajaguru, and T.~Wrase, ``{Automated consistent
  truncations and stability of flux compactifications},''
  \href{http://dx.doi.org/10.1007/JHEP12(2022)026}{{\em JHEP} {\bfseries 12}
  (2022) 026}, \href{http://arxiv.org/abs/2209.08015}{{\ttfamily
  arXiv:2209.08015 [hep-th]}}.

\bibitem{Dibitetto:2019odu}
G.~Dibitetto, J.~J. Fern\'andez-Melgarejo, and M.~Nozawa, ``{6D (1,1) Gauged
  Supergravities from Orientifold Compactifications},''
  \href{http://dx.doi.org/10.1007/JHEP05(2020)015}{{\em JHEP} {\bfseries 05}
  (2020) 015}, \href{http://arxiv.org/abs/1912.04142}{{\ttfamily
  arXiv:1912.04142 [hep-th]}}.

\bibitem{Dibitetto:2010rg}
G.~Dibitetto, R.~Linares, and D.~Roest, ``{Flux Compactifications, Gauge
  Algebras and De Sitter},''
  \href{http://dx.doi.org/10.1016/j.physletb.2010.03.074}{{\em Phys. Lett. B}
  {\bfseries 688} (2010) 96--100},
  \href{http://arxiv.org/abs/1001.3982}{{\ttfamily arXiv:1001.3982 [hep-th]}}.

\bibitem{Dibitetto:2011gm}
G.~Dibitetto, A.~Guarino, and D.~Roest, ``{Charting the landscape of N=4 flux
  compactifications},'' \href{http://dx.doi.org/10.1007/JHEP03(2011)137}{{\em
  JHEP} {\bfseries 03} (2011) 137},
  \href{http://arxiv.org/abs/1102.0239}{{\ttfamily arXiv:1102.0239 [hep-th]}}.

\bibitem{Schon:2006kz}
J.~Schon and M.~Weidner, ``{Gauged N=4 supergravities},''
  \href{http://dx.doi.org/10.1088/1126-6708/2006/05/034}{{\em JHEP} {\bfseries
  05} (2006) 034}, \href{http://arxiv.org/abs/hep-th/0602024}{{\ttfamily
  arXiv:hep-th/0602024}}.

\bibitem{deWit:2002vt}
B.~de~Wit, H.~Samtleben, and M.~Trigiante, ``{On Lagrangians and gaugings of
  maximal supergravities},''
  \href{http://dx.doi.org/10.1016/S0550-3213(03)00059-2}{{\em Nucl. Phys. B}
  {\bfseries 655} (2003) 93--126},
  \href{http://arxiv.org/abs/hep-th/0212239}{{\ttfamily arXiv:hep-th/0212239}}.

\bibitem{Green:1984sg}
M.~B. Green and J.~H. Schwarz, ``{Anomaly Cancellation in Supersymmetric D=10
  Gauge Theory and Superstring Theory},''
  \href{http://dx.doi.org/10.1016/0370-2693(84)91565-X}{{\em Phys. Lett. B}
  {\bfseries 149} (1984) 117--122}.

\bibitem{Ortin:2015hya}
T.~Ortin, \href{http://dx.doi.org/10.1017/CBO9781139019750}{{\em {Gravity and
  Strings}}}.
\newblock Cambridge Monographs on Mathematical Physics. Cambridge University
  Press, 2nd ed.~ed., 7, 2015.

\bibitem{Bergman:2001rp}
O.~Bergman, E.~G. Gimon, and S.~Sugimoto, ``{Orientifolds, RR torsion, and K
  theory},'' \href{http://dx.doi.org/10.1088/1126-6708/2001/05/047}{{\em JHEP}
  {\bfseries 05} (2001) 047},
  \href{http://arxiv.org/abs/hep-th/0103183}{{\ttfamily arXiv:hep-th/0103183}}.

\bibitem{Gimon:1996rq}
E.~G. Gimon and J.~Polchinski, ``{Consistency conditions for orientifolds and
  D-manifolds},'' \href{http://dx.doi.org/10.1103/PhysRevD.54.1667}{{\em Phys.
  Rev. D} {\bfseries 54} (1996) 1667--1676},
  \href{http://arxiv.org/abs/hep-th/9601038}{{\ttfamily arXiv:hep-th/9601038}}.

\bibitem{Kaloper:1999yr}
N.~Kaloper and R.~C. Myers, ``{The Odd story of massive supergravity},''
  \href{http://dx.doi.org/10.1088/1126-6708/1999/05/010}{{\em JHEP} {\bfseries
  05} (1999) 010}, \href{http://arxiv.org/abs/hep-th/9901045}{{\ttfamily
  arXiv:hep-th/9901045}}.

\bibitem{Ortin:2020xdm}
T.~Ortin, ``{O(n, n) invariance and Wald entropy formula in the Heterotic
  Superstring effective action at first order in $\alpha'$},''
  \href{http://dx.doi.org/10.1007/JHEP01(2021)187}{{\em JHEP} {\bfseries 01}
  (2021) 187}, \href{http://arxiv.org/abs/2005.14618}{{\ttfamily
  arXiv:2005.14618 [hep-th]}}.

\bibitem{Aldazabal:2008zza}
G.~Aldazabal, P.~G. Camara, and J.~A. Rosabal, ``{Flux algebra, Bianchi
  identities and Freed-Witten anomalies in F-theory compactifications},''
  \href{http://dx.doi.org/10.1016/j.nuclphysb.2009.01.006}{{\em Nucl. Phys. B}
  {\bfseries 814} (2009) 21--52},
  \href{http://arxiv.org/abs/0811.2900}{{\ttfamily arXiv:0811.2900 [hep-th]}}.

\bibitem{Escobar:2018tiu}
D.~Escobar, F.~Marchesano, and W.~Staessens, ``{Type IIA Flux Vacua with Mobile
  D6-branes},'' \href{http://dx.doi.org/10.1007/JHEP01(2019)096}{{\em JHEP}
  {\bfseries 01} (2019) 096}, \href{http://arxiv.org/abs/1811.09282}{{\ttfamily
  arXiv:1811.09282 [hep-th]}}.

\bibitem{EscobarAtienzar:2019zxp}
D.~Escobar~Atienzar, {\em {Type IIA flux vacua with mobile D6-branes and
  \ensuremath{\alpha} '-corrections}}.
\newblock PhD thesis, U. Autonoma, Madrid (main), Madrid, Autonoma U., 3, 2019.

\bibitem{Herraez:2018vae}
A.~Herraez, L.~E. Ibanez, F.~Marchesano, and G.~Zoccarato, ``{The Type IIA Flux
  Potential, 4-forms and Freed-Witten anomalies},''
  \href{http://dx.doi.org/10.1007/JHEP09(2018)018}{{\em JHEP} {\bfseries 09}
  (2018) 018}, \href{http://arxiv.org/abs/1802.05771}{{\ttfamily
  arXiv:1802.05771 [hep-th]}}.

\bibitem{Bergshoeff:2001pv}
E.~Bergshoeff, R.~Kallosh, T.~Ortin, D.~Roest, and A.~Van~Proeyen, ``{New
  formulations of D = 10 supersymmetry and D8 - O8 domain walls},''
  \href{http://dx.doi.org/10.1088/0264-9381/18/17/303}{{\em Class. Quant.
  Grav.} {\bfseries 18} (2001) 3359--3382},
  \href{http://arxiv.org/abs/hep-th/0103233}{{\ttfamily arXiv:hep-th/0103233}}.

\bibitem{Hertzberg:2007wc}
M.~P. Hertzberg, S.~Kachru, W.~Taylor, and M.~Tegmark, ``{Inflationary
  Constraints on Type IIA String Theory},''
  \href{http://dx.doi.org/10.1088/1126-6708/2007/12/095}{{\em JHEP} {\bfseries
  12} (2007) 095}, \href{http://arxiv.org/abs/0711.2512}{{\ttfamily
  arXiv:0711.2512 [hep-th]}}.

\bibitem{Scherk:1979zr}
J.~Scherk and J.~H. Schwarz, ``{How to Get Masses from Extra Dimensions},''
  \href{http://dx.doi.org/10.1016/0550-3213(79)90592-3}{{\em Nucl. Phys. B}
  {\bfseries 153} (1979) 61--88}.

\bibitem{Myers:1999ps}
R.~C. Myers, ``{Dielectric branes},''
  \href{http://dx.doi.org/10.1088/1126-6708/1999/12/022}{{\em JHEP} {\bfseries
  12} (1999) 022}, \href{http://arxiv.org/abs/hep-th/9910053}{{\ttfamily
  arXiv:hep-th/9910053}}.

\bibitem{Martucci:2005rb}
L.~Martucci, J.~Rosseel, D.~Van~den Bleeken, and A.~Van~Proeyen, ``{Dirac
  actions for D-branes on backgrounds with fluxes},''
  \href{http://dx.doi.org/10.1088/0264-9381/22/13/014}{{\em Class. Quant.
  Grav.} {\bfseries 22} (2005) 2745--2764},
  \href{http://arxiv.org/abs/hep-th/0504041}{{\ttfamily arXiv:hep-th/0504041}}.

\bibitem{Choi:2018fqw}
J.~Choi, J.~J. Fern\'andez-Melgarejo, and S.~Sugimoto, ``{Deformation of $
  \mathcal{N} $ = 4 SYM with varying couplings via fluxes and intersecting
  branes},'' \href{http://dx.doi.org/10.1007/JHEP03(2018)128}{{\em JHEP}
  {\bfseries 03} (2018) 128}, \href{http://arxiv.org/abs/1801.09394}{{\ttfamily
  arXiv:1801.09394 [hep-th]}}.

\end{thebibliography}\endgroup
\end{document}